\documentclass[12pt]{article}
\pdfoutput=1
\usepackage{amssymb, amsmath,amsfonts}

\usepackage{multirow}
\usepackage{mathrsfs}
\usepackage{array}
\usepackage{cite}
\usepackage{booktabs}
\usepackage{tikz}
\usepackage{pgfplots}
\usepackage{float}
\usepackage{subfigure, graphicx}
\usepackage[pdftex, bookmarks=true,colorlinks,linkcolor=red,urlcolor=blue,citecolor=blue]{hyperref}
\usepackage{cite}
\usepackage{slashed}
\usepackage{tabularx}
\usepackage{tikz}
\usepackage{graphics}

\makeatletter\@addtoreset{equation}{section}\makeatother

\def\be{\begin{equation}}
\def\ee{\end{equation}}
\def\bea{\begin{eqnarray}}
\def\eea{\end{eqnarray}}

\def\Dslash{\,\,{\raise.15ex\hbox{/}\mkern-12mu D}}
\def\Dbarslash{\,\,{\raise.15ex\hbox{/}\mkern-12mu {\bar D}}}
\def\delslash{\,\,{\raise.15ex\hbox{/}\mkern-9mu \partial}}
\def\delbarslash{\,\,{\raise.15ex\hbox{/}\mkern-9mu {\bar\partial}}}
\def\pslash{\,\,{\raise.15ex\hbox{/}\mkern-9mu p}}
\def\calDslash{\,\,{\raise.15ex\hbox{/}\mkern-12mu {\cal D}}}

\newcommand{\dd}{\mathrm{d}}

\makeatletter\@addtoreset{equation}{section}\makeatother

\renewcommand{\title}[1]{\vbox{\center\LARGE{#1}}\vspace{5mm}}
\renewcommand{\author}[1]{\vbox{\center#1}\vspace{5mm}}
\newcommand{\address}[1]{\vbox{\center\em#1}}

\def\arXiv#1{\href{http://arxiv.org/abs/#1}{arXiv:#1}}
\def\arXiv#1#2{\href{http://arxiv.org/abs/#1}{arXiv:#1}}

\def\doi#1#2{\href{http://doi.org/#1}{#2}}

\newcommand{\abs}[1]{\left| #1 \right|}

\usepackage{pgfplots}
\pgfplotsset{compat=1.18}

\begin{document}

\unitlength = .8mm

\begin{titlepage}
\vspace{.5cm}
%\preprint{} {\color{violet}{\large{\today}}}
 
\begin{center}
\hfill \\
\hfill \\
\vskip 1cm

\title{\boldmath Analytical studies on 3D hairy rotating\\ black hole interiors 
}
\vskip 0.5cm
{Ling-Long Gao$^{\,a,b}$}\footnote{Email: {\tt linglonggao@buaa.edu.cn}},
{Yan Liu$^{\,a,b}$}\footnote{Email: {\tt yanliu@buaa.edu.cn}} and
{Rui-Long Zhao$^{\,a}$}
%\footnote{Email: {\tt XXX}}
%\blfootnote{*Corresponding author.} 

\address{${}^{a}$Center for Gravitational Physics, Department of Space Science,\\ %and International Research Institute of Multidisciplinary Science, \\
Beihang University, Beijing 100191, China}

\address{${}^{b}$Peng Huanwu Collaborative Center for Research and Education, \\Beihang University, Beijing 100191, China}

\end{center}
\vskip 1.5cm

\abstract{ 
We present an analytical study of the interior  structure of hairy rotating black holes in three-dimensional Einstein gravity,  minimally coupled to a complex  scalar field with a super-exponential potential. 
The interior dynamics of these black holes are characterized by an infinite sequence of Kasner epochs, separated by inversion and  transitions, each of which admits an  analytical description. We derive an explicit analytical expression for this infinite sequence of epochs. At late interior times, the geometry evolves into a curvature singularity, despite the local resemblance of each Kasner epoch to a regular Milne universe on a circle. These results reveal an interior structure richer and more complex than that of its 4D static black hole counterparts.
}
\vfill

\end{titlepage}

\begingroup 
\hypersetup{linkcolor=black}
\tableofcontents
\endgroup

%%%%%%%%%%%%%%%%%%%%%%%%%%%%%%%%%%%%%%%%%

%\newpage
%\vspace{2cm}
%%%%%%%%%%%%%%%%%%%%%%%%%%%%%%%%%%%%%%%%%

%%%%%%%%%%%%% TITLEPAGE %%%%%%%%%%%%%%

%\eject \tableofcontents%

%%%%%%%%%%%%%%%%%%%%%%%%%%%%%%
%%%%%%%%%%%%%%%%%%%%%%%%%%%%%%
%%%%%%%%%%%%%%%%%%%%%%%%%%%%%
\setcounter{footnote}{0}
\section{Introduction}

Three-dimensional black holes provide a mathematically tractable framework  for exploring fundamental questions in quantum gravity. In three dimensions, pure Einstein gravity is topological and propagates no local dynamical degrees of freedom.  Nevertheless, the BTZ black hole \cite{Banados:1992gq} retains key features of  higher-dimensional black holes, such as an event horizon, Hawking temperature and Bekenstein-Hawking entropy. This unique combination makes BTZ black hole  an ideal testing ground for the AdS/CFT correspondence (see e.g. \cite{Li:2008dq,Liu:2009bk,Castro:2011zq}). 

When additional matter fields are included, three-dimensional gravity supports a wider range of black hole solutions.
These solutions, unlike the vacuum BTZ case, possess local dynamical degrees of freedom and typically develop curvature singularities in their interiors. Understanding these interior structures are believed to be crucial for advancing holography, though the precise holographic interpretation of the region behind the horizon remains elusive. 

In higher dimensions, the interiors of static AdS–Schwarzschild black holes are described by a single Kasner geometry with fixed exponents. 
Various deformations to this model can alter these exponents \cite{Frenkel:2020ysx}. Consequently, black hole interiors, ranging from static to stationary configurations, have been extensively studied (see e.g. \cite{Gao:2023rqc, Hartnoll:2022rdv, Hartnoll:2020rwq,
%Hartnoll:2020fhc,
Arean:2024pzo, Caceres:2023zhl,
Liu:2021hap, 
Liu:2022rsy, 
Gao:2023zbd, Sword:2022oyg,
Wang:2020nkd, 
An:2022lvo, Cai:2023igv, Prihadi:2025czn, Carballo:2024hem, Caceres:2024edr, Oling:2024vmq, Zhang:2025hkb
}). These developments are the AdS generalizations of the BKL  dynamics. Achieving a fully analytical description of such interiors remains a central open challenge, one that is essential for constructing a complete  holographic dictionary that includes the interior spacetime of black holes.
%. Such a description is indispensable, particularly for our ambition to construct a complete holographic dictionary that encompasses the interior spacetime of black holes.

{\color{black} Our work is strongly motivated by the need to analytically understand the interior structure and singularity behavior of rotating (stationary) black holes --- a natural and timely extension beyond the now well-studied static cases in the literature \cite{Frenkel:2020ysx,  Hartnoll:2022rdv, Hartnoll:2020rwq,
%Hartnoll:2020fhc,
Arean:2024pzo, Caceres:2023zhl,
Liu:2021hap, 
Liu:2022rsy, 
Gao:2023zbd, Sword:2022oyg,
Wang:2020nkd, 
An:2022lvo, Cai:2023igv, Prihadi:2025czn, Carballo:2024hem, Caceres:2024edr, Oling:2024vmq, Zhang:2025hkb
}. While our previous study \cite{Gao:2023rqc} provided a first numerical exploration in 3D rotating geometries, this work aims to significantly advance  this direction by tackling more general stationary configurations, revealing how rotation fundamentally alters interior dynamics. 
%Analytical solutions are particularly valuable here: (1) They provide greater predictive power than numerical solutions for deep interior regions; (2) holographic detection of singularities urgently requires such exact analytical insights. Our study provides these, deepening our understanding of stationary black hole interiors.
}

The paper is organized as follows. In Sec. \ref{sec:model}, we introduce our theoretical setup. 
In Sec. \ref{sec:kti}, we {\color{black} start from a broad class of scalar potentials, motivated by the  request of rich interior structure with Kanser transitions and asymptotically AdS boundary. Then we select a particular potential with specific parameters as a representative example} and present its numerical solution for black hole interiors. These solutions exhibit two  different typical behaviors for the scalar field ``velocity'' $v$ (defined in 
\eqref{trans-v}), as shown in Fig.  \ref{fig-inv-tran} and Fig. \ref{fig-no-inv-tran}. The spacetime evolves from horizon through an infinite sequence of Kasner epochs, either with or without a  Kasner inversion. We then derive  analytical expressions (Eqs. \eqref{expr-v-T} and \eqref{expr-T}) describing the bounce between two  neighboring epochs. Unlike its 4D counterpart, where neglecting some terms during the bounce leads to monotonically decreasing  $|v|$, the 3D dynamics allows $|v|$ between two  neighboring epochs to either decrease or increase monotonically, even inside non-rotating black holes in principle. Inside rotating black holes $|v|$ between two  neighboring epochs always increases at late times, although $|v|$ could decrease at earlier times if an inversion occurs. It turns out that different behaviors of $v$ are linked to the nature of the final singularity.

In Sec. \ref{sec:Late-time evolution}, we derive three recurrence relations (Eqs. \eqref{eq-rela1-abs}, \eqref{eq-rela2} and \eqref{eq-rela3}), from which the late-time interior evolution can be analytically solved by approximately treating the index number $n$ of Kasner epoch as a continuous parameter. At extremely late interior times, the geometry  evolves towards a special Kasner singularity with $p_t=1$ and $p_x=0$. It seems that such a Kasner metric describes a regular spacetime, however, the way that the interior approaches the Kasner metric indicates that it remains a curvature singularity. If an inversion occurs at a certain finite interior time, the interior could evolve towards $p_t=0$ and $p_x=1$ before the inversion, provided that the number of Kasner transitions is sufficiently large. This metric describes a causal singularity in non-rotating BTZ black holes, however, the interior is again proven to remain a curvature singularity. 

%%%%%%%%%%%%%%%%%%%%%%%%%%%%%%%%%%%%
\section{Setup of the model}
\label{sec:model}
%%%%%%%%%%%%%%%%%%%%%%%%%%%%%%%%%%%%

We consider three dimensional gravity coupled to a complex scalar field with the following action
\begin{align}
\label{eq-action}
    S = \int \dd^3x \sqrt{-g} \left( R -  \partial_a \varphi \partial^a \varphi^* -V(\varphi,\varphi^*)\right)\,.
\end{align}
We have set $ 16\pi G = 1 $ and the AdS radius $L=1$ for convenience. The cosmological constant is included in the potential. The corresponding equations of motion are
\begin{equation}
		\begin{aligned}
		\label{eq-eom}
			R_{ab} - \frac{1}{2} R g_{ab}
			&=\frac{1}{2} \big{(}\partial_a \varphi \partial_b \varphi^* + \partial_a \varphi^* \partial_b \varphi
			 - g_{ab} \left( \partial_c\varphi \partial^c \varphi^* + V\right)\big{)}\,,\\
			\nabla_a \nabla^a \varphi&= \frac{\dd V}{\dd\varphi^* }\,.
		\end{aligned}
\end{equation}
The ansatz of hairy rotating black hole is given by\footnote{Note that the wave-number $n$ on the phase of $\varphi$ was used in \cite{Gao:2023rqc}. Here we use $k$ instead.}
\begin{equation}
\label{eq-ansatz}
		\dd s^2 = \frac{1}{z^2} \left( -f e^{-\chi} \dd t^2 + \frac{\dd z^2}{f} + \left( N \dd t + \dd x \right)^2 \right),
		\quad
		\varphi = \phi(z) e^{-i\omega t + i k x}.
\end{equation}
Here $f,\chi,N,\phi$ are functions of $z$. 
The AdS boundary is located at $z\to 0$. The horizon $z=z_H$ is determined by $f(z_H)=0$. When we further increase the interior time $z>z_H$, the spacetime evolves toward the singularity.  

Substituting the ansatz \eqref{eq-ansatz} into \eqref{eq-eom}, we obtain the equations of motion
	\begin{equation}
		\begin{split}
		\label{eq-full}
		z^3e^{\chi /2}\left ( \frac{fe^{-\chi /2} \phi '}{z}  \right )'& =\phi\left ( k^2z^2-\frac{e^\chi }{f} z^2(\omega +k N)^2 \right )+\frac{1}{2}\frac{\dd V}{\dd\phi} \,,\\
		z e^{-\chi /2}\left ( \frac{e^{\chi /2} N'}{z}  \right ) '&=\frac{2k\phi^2}{f} (\omega +k N)\,,\\
		\frac{\chi '}{2z}-\phi '^2&=\frac{e^\chi \phi ^2}{f^2}(\omega +k N)^2 \, ,\\
		2 e^{\chi /2}z^3\left ( \frac{e^{-\chi /2}f}{z^2}  \right )'&=2k^2z^2\phi ^2+z^2e^\chi N'^2+2V \,.
	\end{split}
	\end{equation}
Here primes denote derivatives with respect to $z$. 
%The co-effect of rotation and complex scalar field could trigger a Kasner inversion, while a super-exponential potential could trigger many Kasner transitions.

%xxxxxxxxxxxxxxxxxxxxxxxxxxxxxxxxxxxxxxxxxxxxx
\section{Kasner inversion and transitions}
\label{sec:kti}

The internal structure of rotating black holes with a scalar potential $V=m^2\varphi^*\varphi$ was studied in \cite{Gao:2023rqc}, revealing both 
the disappearance of the inner horizon and the occurrence of a Kasner inversion. Separately, 
\cite{Hartnoll:2022rdv} investigated planar black hole solutions in 4D gravity coupled to a scalar field with an even  super-exponential potential, reporting the presence of infinitely many Kasner transitions. Here we extend this line of inquiry by studying  rotating black holes in 3D gravity, minimally coupled to a complex scalar field with a super-exponential $V$. 

To ensure that the boundary is asymptotically AdS and the interior undergoes infinitely many Kasner epochs, we choose a scalar potential of the following form,\footnote{{We define the super-exponential term of the full potential as $V_{sup}\equiv k_2 e^{k_3 (\varphi \varphi^*)^{k_4}}$. In this paper we mainly use $V_{sup}$, since the mass-squared term $m^2 \varphi \varphi^*$ and the constant $k_1$ are always negligible compared to $V_{sup}$ deep in the interior.}}
\be 
{
\label{expr-potential-generic}
V(\varphi,\varphi^* )=k_1 +m^2 \varphi\varphi^* + k_2\, e^{k_3(\varphi\varphi^*)^{k_4}}\,,}
\ee
with
\be 
\label{expr-potential-conditions}
-1\le m^2\le 0\,,~~~k_1+k_2=-2\,,~~~k_2>0\,,~~~k_3>0\,,~~~ k_4\in \mathbb{N}^+\,\text{and }k_4\geq 2\, .
\ee
These conditions follow from a perturbative analysis on a single Kasner epoch, in a manner similar to the approach  in \cite{Hartnoll:2022rdv}.
This generic form \eqref{expr-potential-generic} is naturally motivated by the need to have a term in the potential that can dominate at large scalar field modulus $|\varphi|$ to drive Kasner transitions. 
Note that including higher-order polynomial terms in $\varphi\varphi^*$  %or \comment{wrong?} odd functional super-potential terms in $\varphi$ 
does not qualitatively alter the Kasner dynamics, as such contributions become irrelevant within each Kasner epoch.

{In this work we focus on a specific potential without loss of generality, 
	\begin{equation}
	\label{expr-potential}
		V(\varphi,\varphi^* )=-\frac{21}{10}-\frac{3}{4}\,\varphi \varphi ^\ast +\frac{1}{10} \, e^{\frac{1}{10}(\varphi \varphi ^\ast)^2 }\,.
	\end{equation}
We emphasize that the analytical methods used in our work is very generic. The numerical analysis with this specific potential \eqref{expr-potential} only aims to validate the approximations during the analytical calculation. When one use other choices of constants defined in the potential \eqref{expr-potential-generic} under the conditions \eqref{expr-potential-conditions}, one could arrive similar qualitative results, given that the same approximations can be performed. Here we do not aim to scan the parameter space of the potential \eqref{expr-potential-generic} to exhibit their black hole interiors. Instead, we aim to provide powerful analytical methods to study the generic black holes, and uncover non-trivial interior structures.}

{We introduce the coordinate $\rho$ via}
\begin{equation}
{
     \rho \equiv \log z\,.}
\end{equation}
{At the AdS boundary $\rho \rightarrow -\infty$ and at the singularity $\rho \rightarrow +\infty$.
{The ``velocity'' of scalar field is defined as}
\be 
\label{trans-v}
v\equiv\frac{\dd\phi}{\dd \rho}\,.
\ee
At late times the interior will evolve into infinitely many Kasner epochs, of which a single epoch has a constant velocity $v$},
and the metric functions in each Kasner epoch read
\be 
\label{expr-Kas-fchiN}
f\sim -f_K z^{2+v^2}\,,~~~\chi\sim 2v^2\log z+\chi_1\,,~~~N\sim N_K+\frac{E_K}{2-v^2}z^{2-v^2}\,,
\ee 
where $f_K, \chi_1, N_K, E_K$ are constants. {Note that \eqref{expr-Kas-fchiN} with constant $v, f_K, \chi_1, N_K$ and $E_K$ are derived by neglecting all terms at the right sides of the full equations of motion  \eqref{eq-full}, and are thus valid only within a single Kasner epoch, as consistently checked from numerics. The same approximations can be made for any Kasner epoch satisfying \eqref{expr-Kas-fchiN}, with different values of $v, f_K, \chi_1, N_K$ and $E_K$. } 

Performing the coordinate transformation
\be 
\label{tran-tauToz}
\tau=\frac{2}{\sqrt{f_K}(v^2+2)}z^{-(v^2+2)/2}\,,
\ee
the metric and scalar field take the Kasner form
\be 
\label{expr-ds-kas}
\begin{split}
\dd s^2&=-\dd\tau^2+c_t \tau^{2p_t}\dd t^2+ c_x \tau^{2p_x}(N_K \dd t+\dd x)^2\,,\\
\phi&=p_{\phi} \log \tau+c_{\phi}\,,
\end{split}
\ee 
where the Kasner exponents
\be 
\label{expr-kas-exponents}
p_t=\frac{v^2}{v^2+2}\,,~~
p_x=\frac{2}{v^2+2}\,,~~
p_{\phi}=-\frac{2v}{v^2+2}\,,
\ee 
and 
\be
\label{expr-ctcxcphi}
    c_t= f_K e^{-\chi_1}\left( \frac{\sqrt{f_K}(v^2+2)}{2} \right)^{2p_t}, ~~c_x=\left( \frac{\sqrt{f_K}(v^2+2)}{2} \right)^{2p_x},~~c_{\phi}= p_{\phi} \log \left( \frac{\sqrt{f_K}(v^2+2)}{2} \right)\,.
\ee
Since all Kasner exponents are expressed in terms of $v$, we mainly focus on the behavior of $v$ throughout the discussion.

\begin{figure}[h!]
    \centering
    \includegraphics[height=5cm]{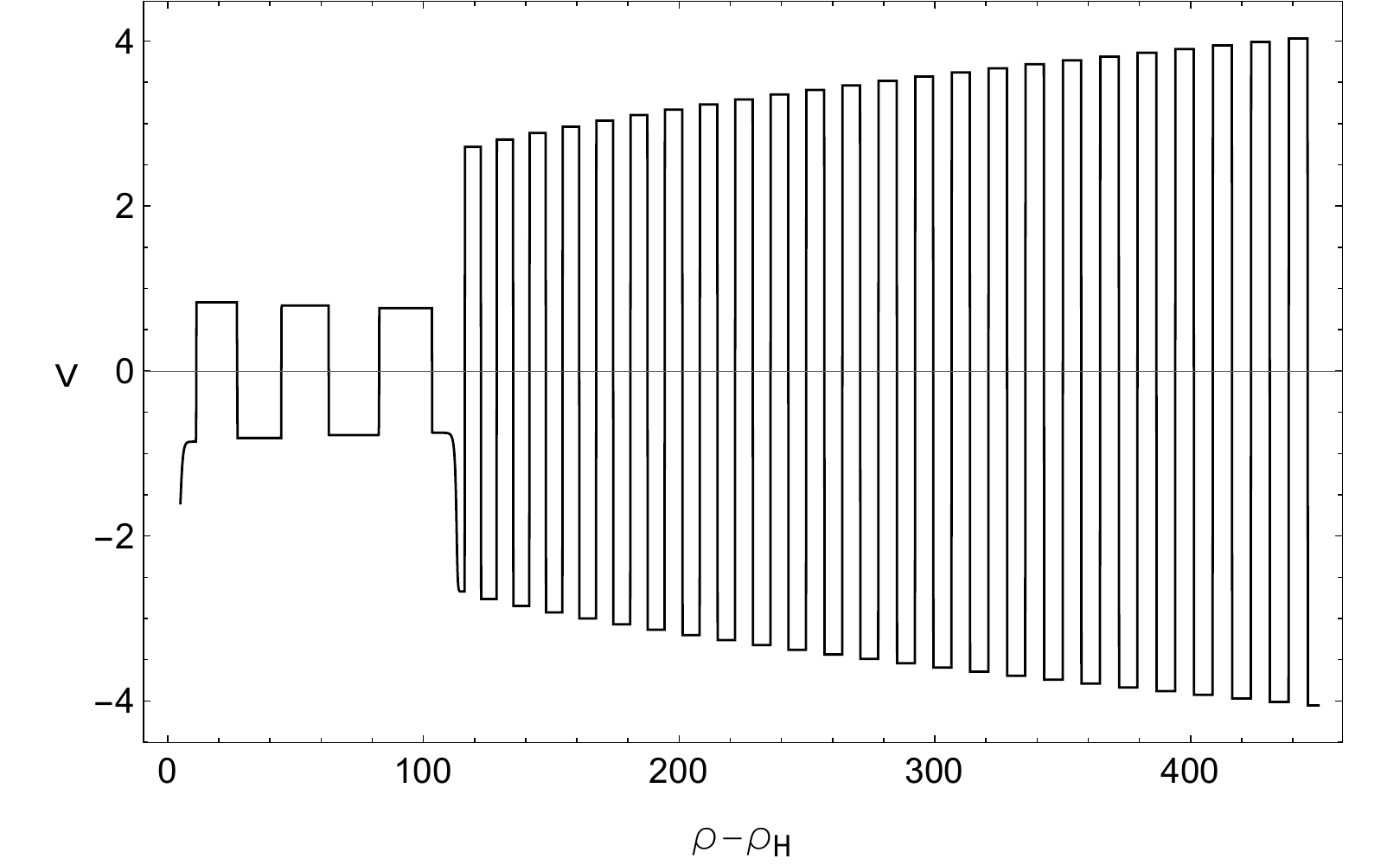}~
    \includegraphics[height=5.1cm]{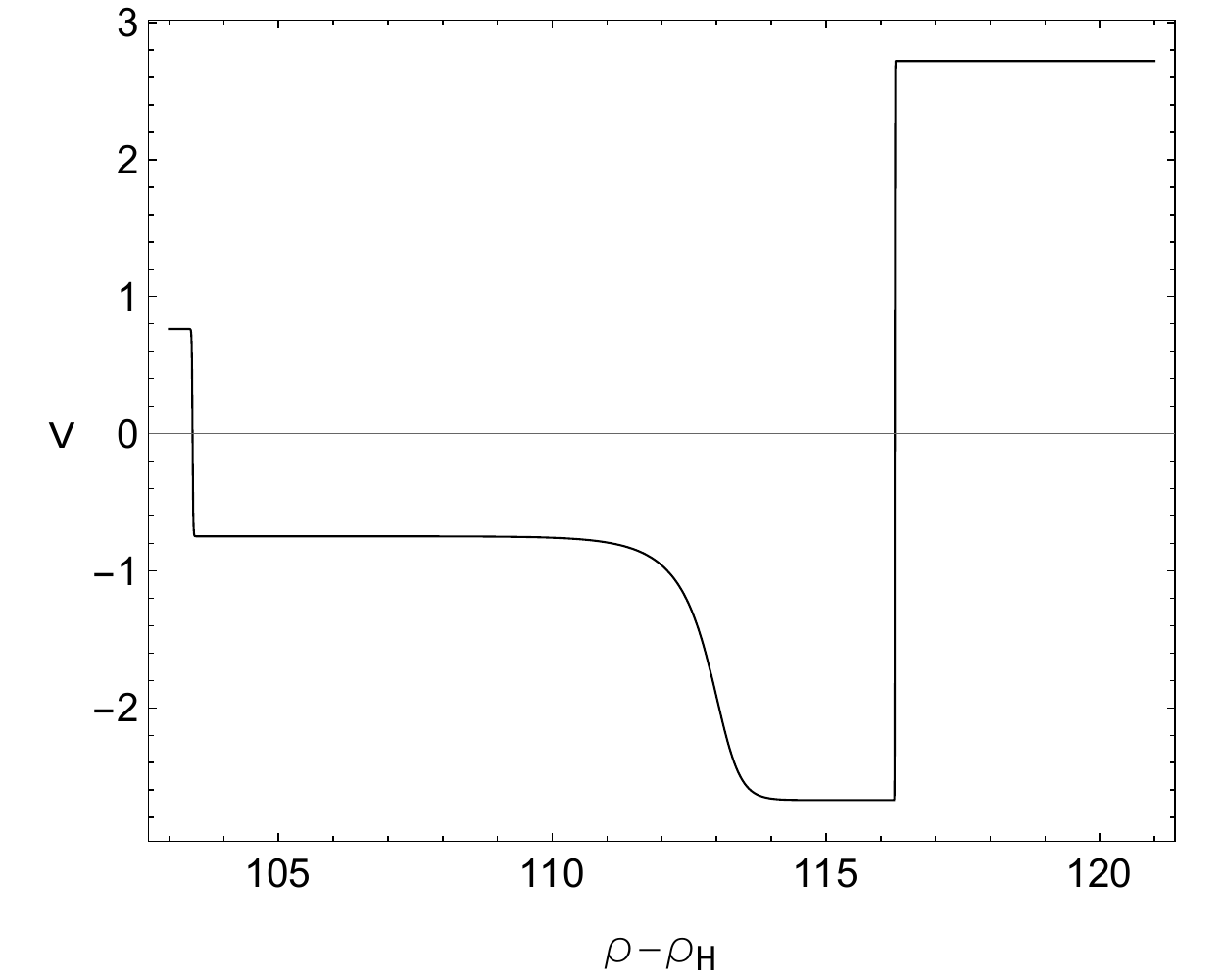}
    \caption{\small Evolution of $v$ as a function of $\rho$. 
We select a specific, yet not finely tuned, set of initial values at the horizon.  
    %A specific but \textcolor{red}{not finely tuned} set of initial values is chosen at the horizon. 
    We name the type of left transitions as ``decreasing'' transitions since $|v|$ between two neighboring epochs decreases constantly, and the type of right ones as ``increasing'' transitions. The inversion separates the two types of transitions. 
    The right plot shows a zoomed-in view of the Kasner inversion region from the left plot.}
    \label{fig-inv-tran}
\end{figure}

\begin{figure}[h!]
    \centering
    \includegraphics[width=0.67\textwidth]{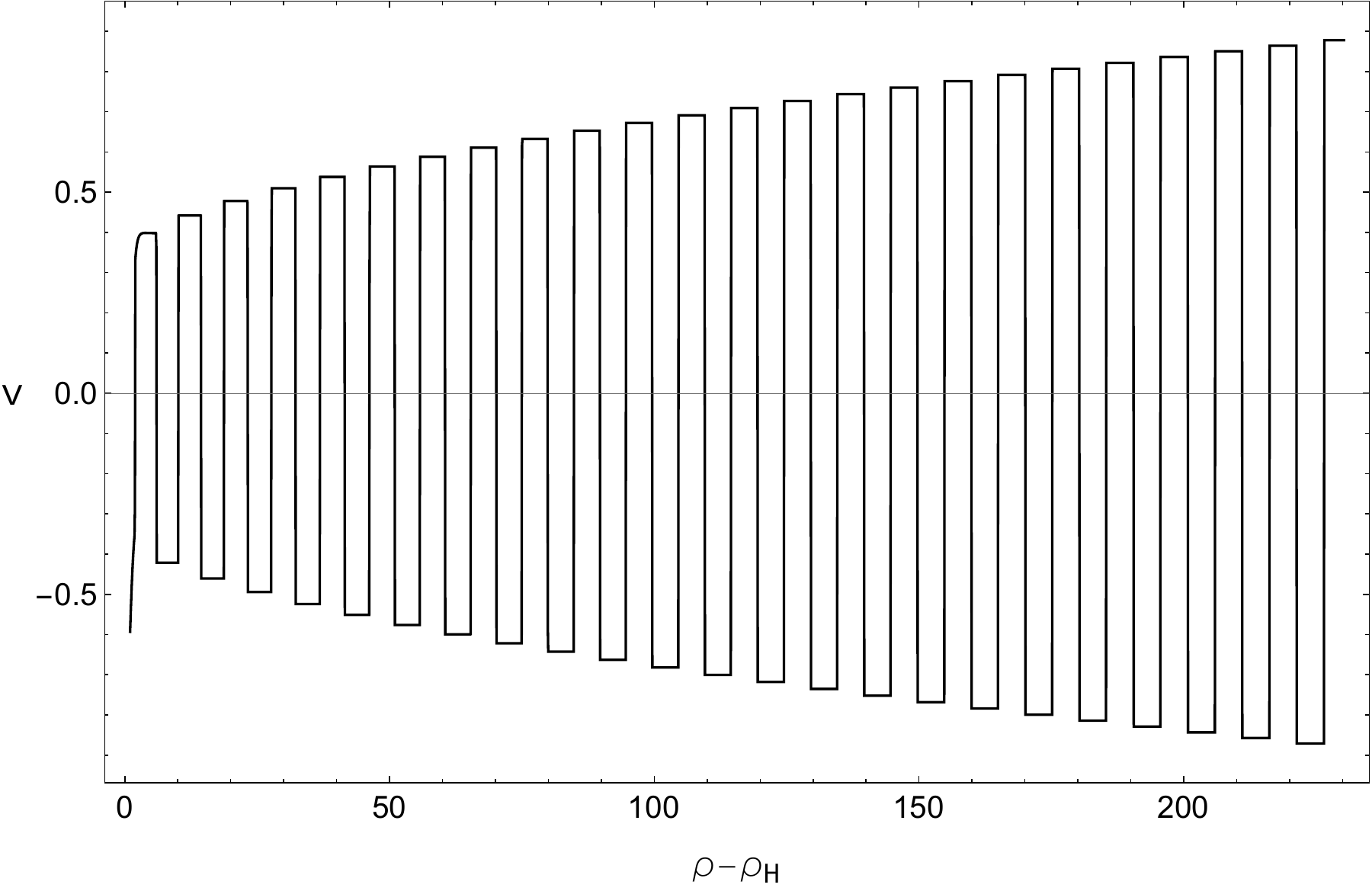} 
    \caption{\small Evolution of $v$ for another different choice of initial values at the horizon. There is no inversion, and only ``increasing'' transitions exist in the interior.}
    \label{fig-no-inv-tran}
\end{figure}

The figures \ref{fig-inv-tran} and \ref{fig-no-inv-tran} show the two types of the interior we have found, in addition to the disappearance of inner horizon previously found in \cite{Gao:2023rqc} for a simple mass-squared term in the potential. {In the left panel of Fig.~\ref{fig-inv-tran} there are two different behaviors of the interiors.} On the left there are some ``decreasing'' transitions where $|v|$ between two neighboring epochs decreases toward the deeper interior, and a Kasner inversion occurs. Then the interior evolves into infinitely many ``increasing'' transitions where $|v|$  between two neighboring epochs increases as $\rho$ increases. The right panel in Fig.~\ref{fig-inv-tran}   focuses on the Kasner  inversion region. 
While in Fig.~\ref{fig-no-inv-tran} there is no Kasner inversion and only increasing transitions exist. {These behaviors are different from the ones in \cite{Hartnoll:2022rdv} where $|v|$ between two  neighboring epochs monotonically decreases as $\rho$ increases.} 

{Starting from the horizon $\rho_h$ to the singularity $\rho \to \infty$, a timelike geodesic satisfying  $g_{tt}(dt/d\tau)^2+g_{\rho\rho}(d\rho /d\tau)^2+g_{xx}(dx/d\tau)^2+2g_{tx}(dt/d\tau)(dx/d\tau)=-1$ with zero conserved `energy' (i.e., $E=-g_{tt}dt/d\tau-g_{tx}dx/d\tau$) and zero angular momentum (i.e.,  $L=g_{xx}dx/d\tau+g_{tx}dt/d\tau$) has a maximal proper time $\tau_m$, which could have a dual description from thermal one-point function \cite{Grinberg:2020fdj}. Here we find that $\tau_m$ is dominated by the contribution from the horizon $\rho_h$ to the initial interior time of  first Kasner epoch $\rho_i$. This indicates that the proper time from the first Kasner epoch to the singularity is extremely small, although there are rich internal structures, for example, an inversion and infinitely many Kasner transitions. Consequently, providing a dual interpretation for the successive Kasner epochs becomes extremely challenging.}
    
    \vskip 0.3cm

{In the following subsections we aim to derive the analytical descriptions of the Kasner inversion and transitions. Within a single Kasner epoch, we have neglected all terms at the right sides of \eqref{eq-full} to get the solution \eqref{expr-Kas-fchiN}. However, for the full interior including the inversion and transitions, the derivative term related to rotation, $z^2 e^\chi N'^2$ is not negligible during the inversion; and the super-exponential potential terms in \eqref{expr-potential-generic},  $V_{sup}=k_2 e^{k_3 (\varphi \varphi^*)^{k_4}}$ and $\dd V_{sup}/\dd \phi\ $ are not negligible during a transition. Therefore, these rotation and potential terms are reconsidered to derive the analytical descriptions, respectively. After the analytical calculation we make some comments on the interior of 4D static black hole counterpart.}

\subsection{Analytical description of Kasner inversion}
\label{sec-inv}

As discussed in Sec. \ref{sec:kti}, within a single Kasner epoch we have neglected all terms at the right sides of the full equations of motion \eqref{eq-full} to derive the solution \eqref{expr-Kas-fchiN}. The rotation term $z^2 e^\chi N'^2$ and super-exponential potential terms $V_{sup}$, $\dd V_{sup}/\dd \phi$ are not negligible when a Kasner inversion or transition occurs. It turns out that the rotation and potential terms trigger the Kasner inversion and transition, respectively. This is consistent with the observation in Fig. \ref{fig-inv-tran} that the Kasner inversion and transitions are independent of each other, i.e.  the region where the rotation term is non-negligible, does not overlap with the region where the potential terms are non-negligible. 

Therefore, in the vicinity of the Kasner inversion, all potential terms can be neglected and the equations of motion \eqref{eq-full} are simplified to
	\begin{equation}
		\begin{split}
		\label{eq-eom9}
			z e^{\chi /2}\left ( \frac{fe^{-\chi /2} \phi '}{z}  \right )'& =0\,,\\
			z e^{-\chi /2}\left ( \frac{e^{\chi /2} N'}{z}  \right ) '&=0\,,\\
			\frac{\chi '}{2z}&=\phi '^2  ,\\
			2 e^{\chi /2}z^3\left ( \frac{e^{-\chi /2}f}{z^2}  \right )' &=z^2e^\chi N'^2\,.
		\end{split}
	\end{equation}
The above simplifications are well supported by our numerical check. Thus we derive a third-order differential equation for the scalar field $\phi$ ,
	\begin{equation}
		\begin{split}
		\label{eq-eom-inv}
    \phi''' - \frac{2\phi''^2}{\phi'} + \left(z\phi'^2 - \frac{3}{z}\right)\phi'' + \phi'^3 - \frac{3\phi'}{z^2} = 0\,,
		\end{split}
	\end{equation}
where $\phi$ is a function of $z$. 
Performing the transformation $z=e^\rho$, the above equation becomes
\be 
\dddot\phi-\frac{2 \ddot\phi^2}{\dot\phi}+\left(\dot\phi ^2-2\right) \ddot\phi=0\,,
\ee 
which can be rewritten in terms of $v$ via $v=\dot{\phi}$,
\be 
\ddot v-\frac{2 \dot v^2}{v}+\left(v ^2-2\right) \dot v=0\,.
\ee 
Here dots denote derivatives with respect to $\rho$. 

The solution is 
\be 
v_{I}=I^{-1}(\rho-\rho_I)
\ee 
with
\be 
\label{expr-vI-inv}
I(x)=-\frac{c_I}{2 \sqrt{c_I^2-8}}\tanh ^{-1}\left[\frac{c_I-2 x}{\sqrt{c_I^2-8}}\right]+\frac{1}{4} \log \left|x^2-c_I x+2\right|-\frac{\log |x|}{2}\,,
\ee
where $c_I$ and $\rho_I$ are two constants of integration. In the parameter region we considered, the function $y=I(x)$ is  one-to-one and onto, and the inverse function $I^{-1}(y)$ is thus well-defined. 
We must have $c_I^2>8$ in order that the analytical solution matches the numerical results. We are only concerned with the two ``plateaus'' of $v_I$, $v_i$ and $v_f$ that correspond precisely to the two values of $x$ at which $I(x)$ diverges. Since the $\tanh^{-1}$ term dominates the divergence in \eqref{expr-vI-inv} over all the other contributions, we work with the approximate inverse function
\be 
I_{appr}(x)=-\frac{c_I}{2 \sqrt{c_I^2-8}}\tanh ^{-1}\left[\frac{c_I-2 x}{\sqrt{c_I^2-8}}\right]\,,
\ee 
which corresponds to the approximate solution for $v$,
\be 
v_{appr}^I=\frac{c_I}{2}+\frac{1}{2} \sqrt{c_I^2-8} \tanh \left[\frac{2 \sqrt{c_I^2-8} }{c_I}\left(\rho -\rho _I\right)\right]\,.
\ee
Thus we have 
\be 
v_n=\lim_{\rho\rightarrow -\infty}v^I_{appr}(\rho)\,,~~~ v_{n+1}=\lim_{\rho\rightarrow +\infty}v^I_{appr}(\rho)\,,
\ee
and more precisely
\be 
\begin{aligned}
v_i=\frac{1}{2} \left(c_I-\sqrt{c_I^2-8}\right)\,,~~~
v_f=\frac{1}{2} \left(c_I+\sqrt{c_I^2-8}\right)\,,
\end{aligned}
\qquad \text{if~$c_I>2\sqrt{2}\,,$}
\ee 
and 
\be
\begin{aligned}
v_i=\frac{1}{2} \left(c_I+\sqrt{c_I^2-8}\right)\,,~~~
v_f=\frac{1}{2} \left(c_I-\sqrt{c_I^2-8}\right)\,,
\end{aligned}
\qquad \text{if~ $c_I<-2\sqrt{2}\,.$}
\ee
The above results can be rewritten in a simpler way,
\be 
|v_i|=\frac{1}{2} \left(|c_I|-\sqrt{c_I^2-8}\right)\,,~~~|v_f|=\frac{1}{2} \left(|c_I|+\sqrt{c_I^2-8}\right)\,,~~~v_i v_f>0\,,
\ee 
which permits at most one Kasner inversion
\be 
\label{expr-v-inver}
{v}\rightarrow {\frac{2}{v}}\,.
\ee 

{From the above transformation rule of $v$ for the inversion, we define the critical value of $v$ as}
\be 
{
v_c^I=\sqrt{2}\,.}
\ee 
%The expression  for $N$ in \eqref{expr-Kas-fchiN} within a Kasner epoch indicates that the mechanism of the Kasner inversion is as follows. 
If $\left|v \right|<v_c^I$ for a certain Kasner epoch, 
from the expression  for $N$ in \eqref{expr-Kas-fchiN} within a Kasner epoch, 
$N'$ will become progressively larger as $\rho$ increases.  Consequently, the rotation terms will become non-negligible and an inversion occurs, until $\left|v \right|$ is abruptly raised above the critical value $v_c^I$ by the rule \eqref{expr-v-inver}. This is precisely the mechanism of the Kasner inversion. After the inversion, $N'$ will decrease continuously, and the stability of a Kasner epoch will no longer be affected by rotation terms, which ensures that no further inversions can occur. 
%. \textcolor{red}{Thus no inversion will occur again.}

\begin{figure}[htbp]
    \centering
    \includegraphics[width=0.46\textwidth]{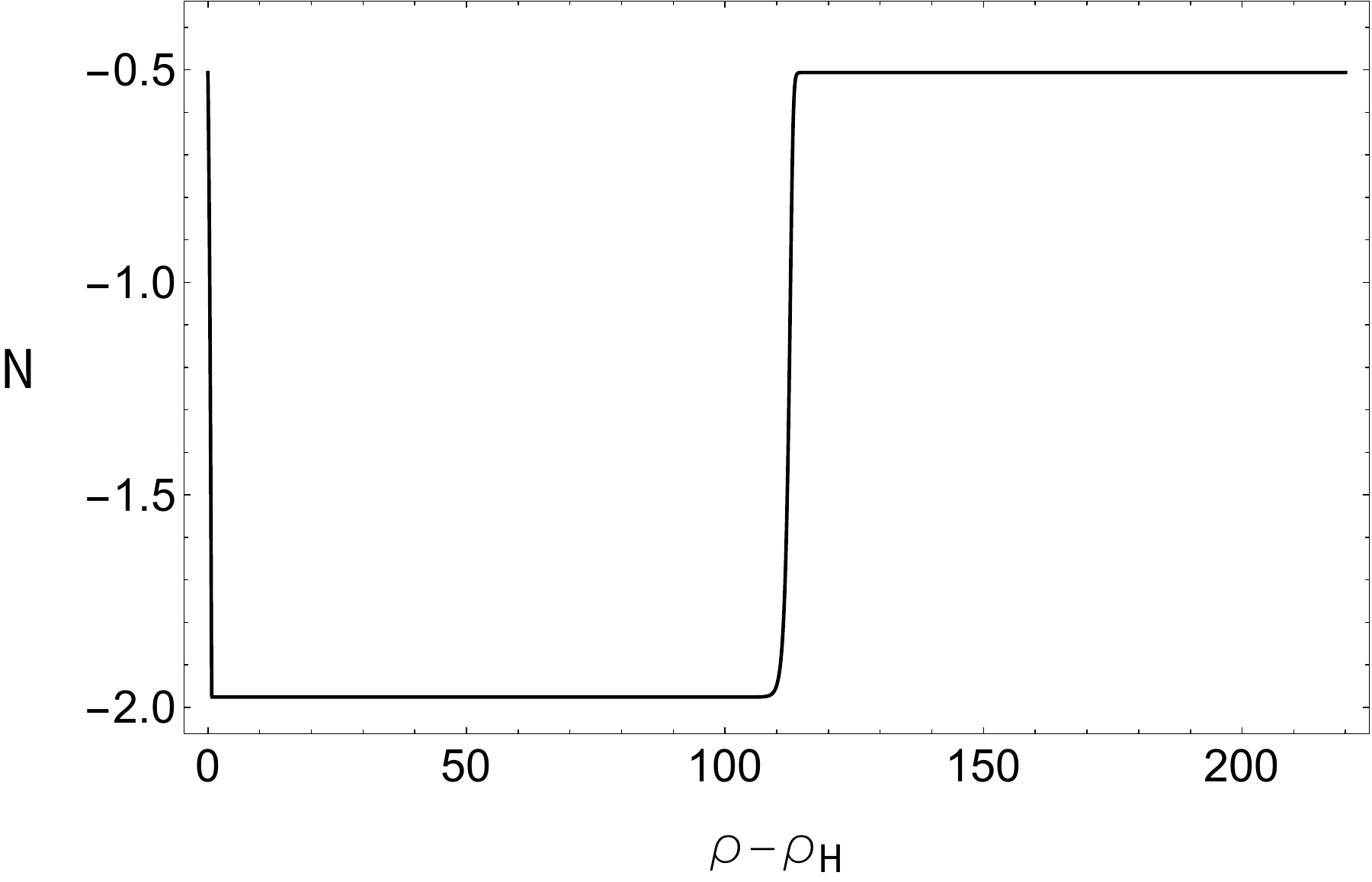}~~~
     \includegraphics[width=0.45\textwidth]{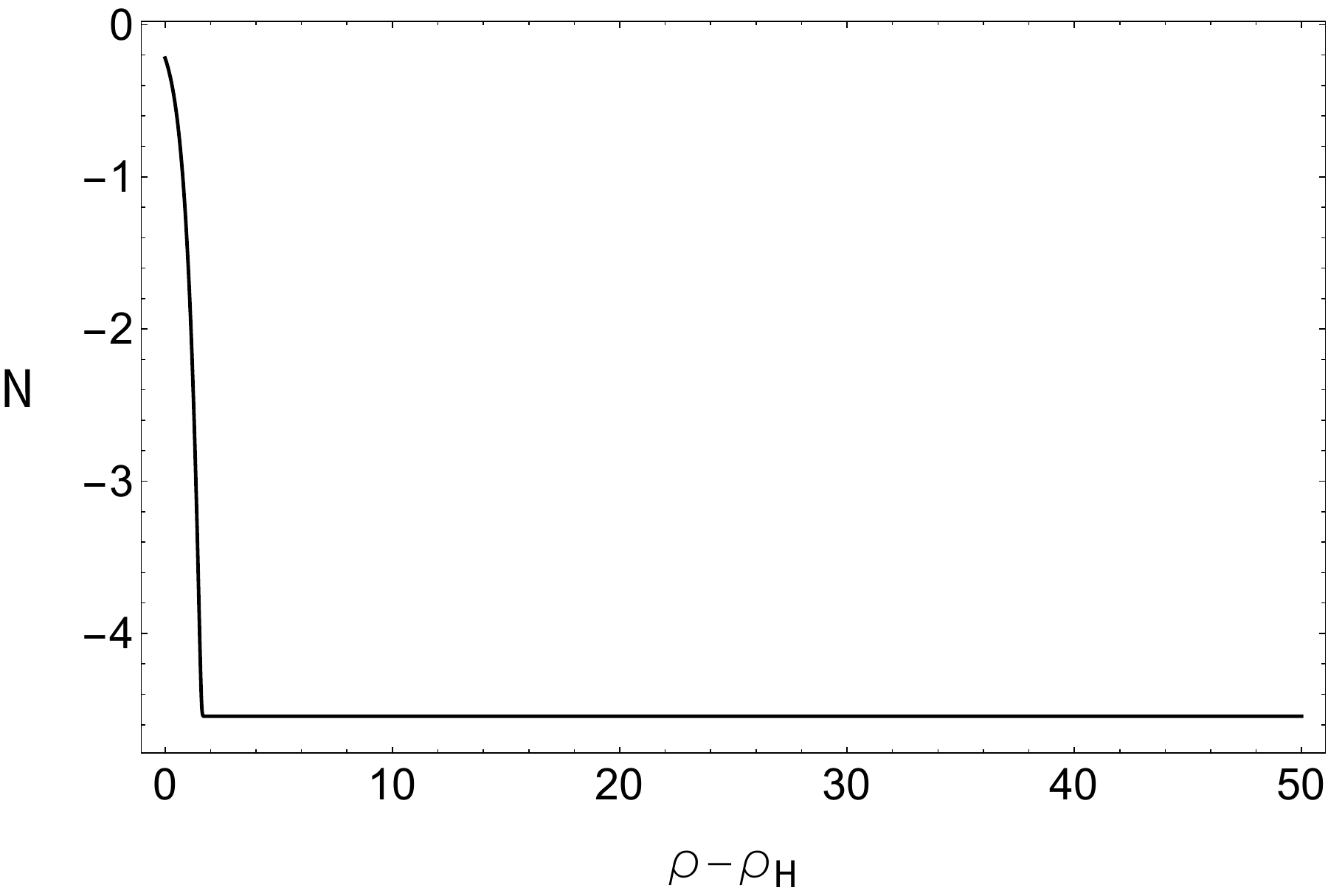}
    \vspace{-0.3cm}
    \caption{\small Evolution of $N$ with ({\em left}) and without ({\em right}) Kasner inversion. In the left, we have $N_h=-0.506$, $N_K^i=-1.976$, and $N_K^f=-0.506$; while in the right, $N_h=-0.220$ and $N_K^f=-4.544$.}
\label{fig-two-N}
\end{figure}

From Fig. \ref{fig-two-N} we observe that these transitions indeed do not affect the behavior of the metric function field $N$, which is consistent with the argument that the inversion and transitions are independent of each other as discussed before.\footnote{We do not show the other extreme case of $N$ in \cite{Gao:2023rqc}, where there is no Kasner epoch before the inversion which leads to that $N$ is not constant before the inversion.} We also confirm the relation of $N$ among these values: $N_h$, the value of $N$ at horizon; $N_K^i$, the value of $N$ before the inversion (if it exists); and $N_K^f$, the value of $N$ at late time (after the inversion). If there is no inversion, 
\be 
N_K^f=1/N_h=-k/\omega\,,
\ee 
while if the inversion occurs, \be N_K^f=1/N_K^i=N_h=-\omega/k\,.\ee {These relations 
were first discovered numerically in our earlier study \cite{Gao:2023rqc}, where a simple mass-squared term is considered in the potential. Here, we numerically confirmed that these relations also hold  for the potential with a super-exponential term. These findings suggest that the relations are  universal in 3D rotating black holes coupled to scalar fields. %, and they must be provable through additional physical constraints beyond the equations of motion.
}

\subsection{Analytical description of Kasner transitions
 }
\label{sect-tran}

As discussed in Sec. \ref{sec-inv}, the Kasner inversion is driven by the rotation term, while the  transitions are triggered by potential terms. In the vicinity of a Kasner transition, the rotation term can thus be neglected. The equations of motion \eqref{eq-full} are simplified to
{
	\begin{equation}
	\label{eq-eom-tran}
			\begin{split}
				z^3e^{\chi /2}\left ( \frac{fe^{-\chi /2} \phi '}{z}  \right )'& =\frac{1}{2}\frac{\dd V_{sup}}{\dd\phi}\,,\\ 
			%	z e^{-\chi /2}\left ( \frac{e^{\chi /2} N'}{z}  \right ) '&=0\,, \\
				\frac{\chi '}{2z}&=\phi '^2 \,, \\
				2 e^{\chi /2}z^3\left ( \frac{e^{-\chi /2}f}{z^2}  \right )' &=2 V_{sup}\,.
			\end{split}
		\end{equation}}
		
{Parameterizing the super-exponential part $V_{sup}$ as 
\begin{equation}
\label{expr-potential-repara}
    V_{sup}=k_2 e^{k_3 (\varphi \varphi^*)^{k_4}}\equiv \, e^{\int H(\phi )\,\dd\phi }\,,
\end{equation}}
we derive a third-order equation for the scalar field,
\be 
\label{eq-phi-tran}
\frac{\dddot{\phi}}{\ddot{\phi}}=\left(H+\frac{H'}{H}\right) \dot{\phi}-2+ \left(\frac{4}{H}-3\dot{\phi}+\frac{2\dot{\phi}^2}{H}\right)\dot{\phi}-\frac{4}{H}\ddot{\phi}+\frac{\dddot{\phi}}{\ddot{\phi}}\frac{2\dot{\phi}}{H}\,,
%\text{additional terms}
\ee 
which can be rewritten in terms of $v$
\be 
\label{eq-tran-v}
\frac{\ddot v}{\dot v}=\left(H+\frac{H'}{H}\right)v-2+\left(\frac{4}{H}-3v+\frac{2v^2}{H}\right)v-\frac{4}{H}\dot v+\frac{\ddot v}{\dot v}\frac{2v}{H}\,.
\ee 
Here the prime denotes a derivative with respect to $\phi$, and dots denote derivatives with respect to $\rho$. %\comment{Compared to 4D}

\begin{figure}[htbp]
    \centering
    \includegraphics[width=0.6\textwidth]{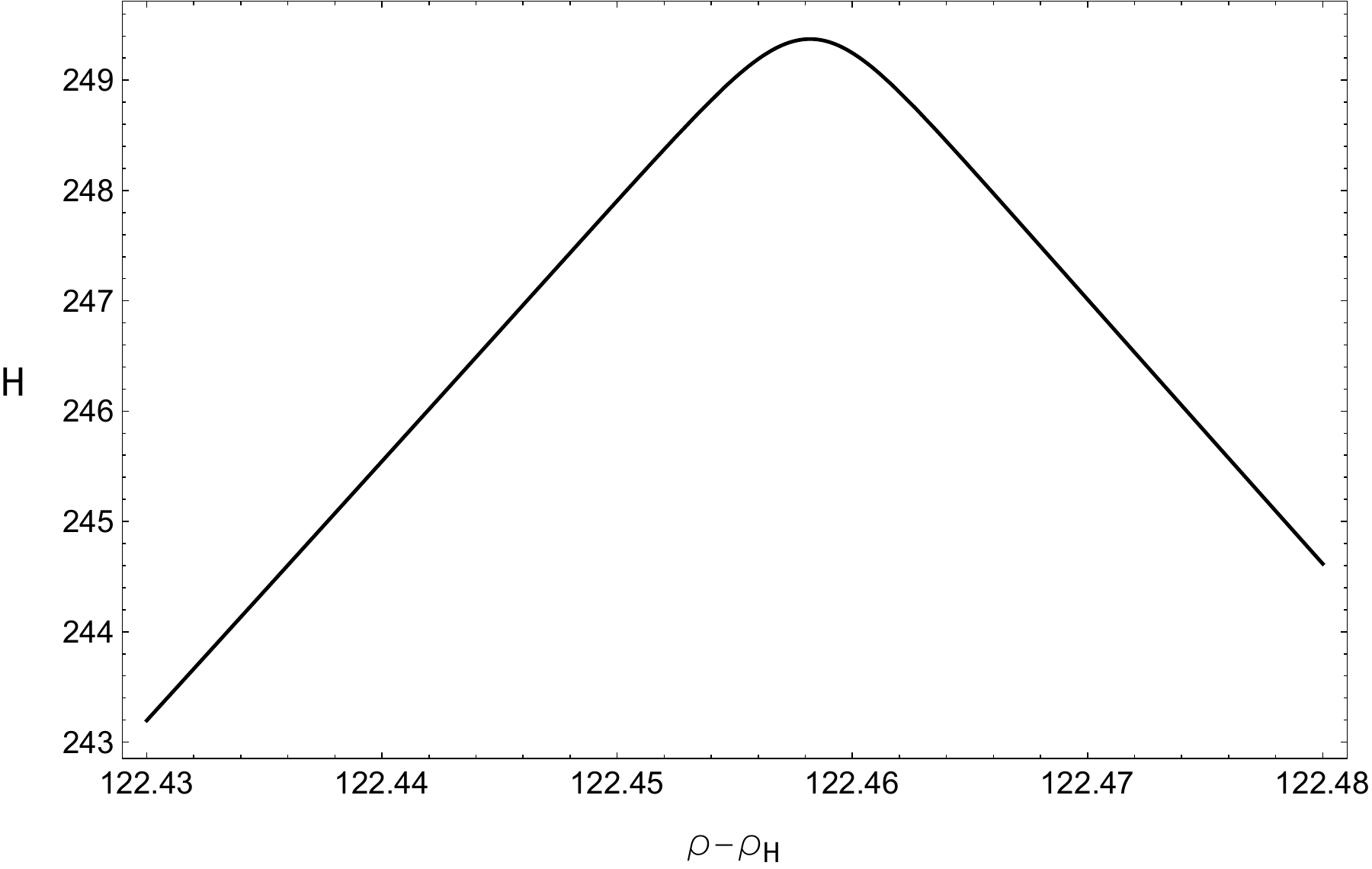}~~~
    \vspace{-0.3cm}
    \caption{\small This figure shows that 
    $H$ {satisfies \eqref{eq:Hproperity} and is %\textcolor{red}{large enough and }
    almost a constant $H_n=2/h_n$} within a selection of an arbitrary $n$-th Kasner transition.
   % We have selected an arbitrary $n$-th Kasner transition and find that $H$ is almost a constant $H_n$.
   }
    \label{fig-H}
\end{figure}

%It is very hard to solve the above equation directly.
Directly solving this equation is formidable. However, we can make an approximation within the $n$-th bounce
\be 
\label{eq:Hproperity}
\abs{\frac{H'}{H}}\ll \abs H\,,~~
H=H_n \equiv \frac{2}{h_n} \,\text{is a constant,}
\ee 
which is supported by Fig. \ref{fig-H}.
Therefore, the equation \eqref{eq-tran-v} is simplified to
\be 
\label{eq-v-simp}
\frac{\ddot v}{\dot v}=\frac{2}{h_n} v-2+(2h_n-3v+h_n v^2)v-2h_n\dot v+h_n \frac{\ddot v}{\dot v}v\,.
\ee 
The corresponding solution is 
%a little complicated
\be
\label{expr-v-T}
v_T=T^{-1}(\rho-\rho_n)\,,
\ee 
with the inverse function
\be 
\label{expr-T}
\begin{split}
T(x)=&-\frac{c_T \left(1-J_n^2\right)}{2 J_n^2 \sqrt{c_T^2-4 J_n^2}}\tanh ^{-1}\left[\frac{c_T+2 y(x)}{\sqrt{c_T^2-4 J_n^2}}\right]-\frac{1-J_n^2}{2J_n^2}\log |y(x)|\\&+\frac{1-J_n^2}{4J_n^2}\log \left|y(x)^2+y(x)c_T+J_n^2\right|
\end{split}
\ee 
%\be 
%\label{expr-T}
%\begin{split}
%T(x)&=\frac{2(-1+J_n^2)\sqrt{c_2^2+4J_n^2}}{4 c_2 J_n^2}\arctanh{(\frac{\sqrt{c_2^2+4J_n^2}-2y(x)}{c_2})}\\&+\frac{-1+J_n^2}{4J_n^2}\left(2 \log {y(x)}-\log{(-y(x)^2+y(x) \sqrt{c_2^2+4J_n^2}-J_n^2)}\right)\,,
%\end{split}
%\ee
where 
\be 
J_n=\sqrt{1-2h_n^2}\,,\quad y(x)=1-h_n x\,,
\ee 
and $c_T$ and $\rho_n$ are two constants of integration. In order that the analytical solution matches the numerical results $c_T$ must satisfy $c_T^2>4J_n^2\,.$ We are only concerned with the two plateaus of $v_T$, $v_n$ and $v_{n+1}$ that correspond to the two values of $x$ at which $T(x)$ diverges. Since the $\tanh^{-1}$ term dominates the divergence over all the other contributions in \eqref{expr-T}, we work with the approximate inverse function 
\be 
T_{appr}(x)=-\frac{c_T \left(1-J_n^2\right)}{2 J_n^2 \sqrt{c_T^2-4 J_n^2}}\tanh ^{-1}\left[\frac{c_T+2 y}{\sqrt{c_T^2-4 J_n^2}}\right]\,,
\ee 
which correspond to the approximate $v$,
\be 
\label{expr-vappr}
v^T_{appr}(\rho)=\frac{c_T+2}{2 h_n}+\frac{\sqrt{c_T^2-4 J_n^2} }{2 h_n}\tanh \left[\frac{2 J_n^2 \sqrt{c_T^2-4 J_n^2}}{c_T \left(1-J_n^2\right)}\left(\rho -\rho _n\right)\right]\,.
\ee 
Thus we have
\be 
v_n=\lim_{\rho\rightarrow -\infty}v^T_{appr}(\rho)\,,~~~ v_{n+1}=\lim_{\rho\rightarrow +\infty}v^T_{appr}(\rho)\,.
\ee
{We can also make a further reasonable assumption,
\be 
\label{expr-assump-hn}
0<h_n<\frac{\sqrt 2}{2}\,,
\ee 
which is again strongly supported by Fig. \ref{fig-H}.}
Thus we have
\be 
\label{expr-vn-ct-neg}
v_n=\frac{c_T+2}{2 h_n}+\frac{\sqrt{c_T^2-4 J_n^2}}{2 h_n}\,,~~~v_{n+1}=\frac{c_T+2}{2 h_n}-\frac{\sqrt{c_T^2-4 J_n^2}}{2 h_n}\,,~~\text{if~$c_T<-2J_n\,,$}
\ee 
or
\be 
\label{expr-vn-ct-pos}
v_n=\frac{c_T+2}{2 h_n}-\frac{\sqrt{c_T^2-4 J_n^2}}{2 h_n}\,,~~~v_{n+1}=\frac{c_T+2}{2 h_n}+\frac{\sqrt{c_T^2-4 J_n^2}}{2 h_n}\,,~~\text{if~$c_T>2J_n\,.$}
\ee 

In the following we focus on the two values,
$\Delta v_n=|v_{n+1}|-|v_n|$ and $s_n=v_n v_{n+1}$. The former determines whether the transitions are of increasing or decreasing type, which corresponds to two completely different behaviors of the singularity at extremely late times, as discussed in the next section. The latter determines whether $v_n$ alternates in sign, i.e., whether $\phi$ oscillates or change monotonically at late times. The behaviors of these two values depend on a critical value of $v$ for transitions, which is defined as
\be
\label{eq:criticalv}
{
v_c^T=\sqrt{2}\,.}
\ee 
{Therefore, this critical value is used to distinguish the increasing and decreasing transitions.}

Firstly, for the solution \eqref{expr-vn-ct-neg}, if we begin with a Kasner epoch with $|v_n|>v_c^T$, then we have
\be 
c_T<-2\,,
\ee 
from which we have 
\be 
\label{expr-dvpos-snneg}
\Delta v_n=-\frac{2+c_T}{|h_n|}>0\,,~~~s_n=-2+\frac{2+c_T}{h_n^2}<0\,.
\ee
\textcolor{black}{Thus $|v_{n+1}|>v_c^T$ still holds and then $\left|v_{n+2}\right|>\left|v_{n+1}\right|.\,$} Therefore, $|v|$ always increases from bounce to bounce. This case corresponds to the increasing transitions where $v$ alternates in sign, as shown in Fig. \ref{fig-no-inv-tran} and the right portion of Fig. \ref{fig-inv-tran}. \textcolor{black}{If we begin with $|v_n|<v_c^T$, then $c_T>-2$ and there are two cases for $s_n$}
\be 
\label{expr-dvneg-snneg}
\Delta v_n=-\frac{2+c_T}{|h_n|}<0\,,~~~\text{if~} s_n=-2+\frac{2+c_T}{h_n^2}<0\,,
\ee 
or
\be 
\label{expr-dvneg-snpos}
\Delta v_n=-\frac{\sqrt{c_T^2-4J_n^2}}{|h_n|}<0\,,~~~\text{if~}s_n=-2+\frac{2+c_T}{h_n^2}>0\,.
\ee 
\textcolor{black}{Thus $|v_{n+1}|<v_c^T$ still holds and then $\left|v_{n+2}\right|<\left|v_{n+1}\right|.\,$} Therefore, $|v|$ always decreases from bounce to bounce. The case \eqref{expr-dvneg-snneg}
corresponds to the decreasing transitions where $v$ alternates in sign, as shown in the left portion in Fig. \ref{fig-inv-tran}. The case \eqref{expr-dvneg-snpos} corresponds to a step-wise decreasing $|v_n|$, however, we have not yet found this type of transitions numerically. \textcolor{black}{Secondly, for the solution \eqref{expr-vn-ct-pos},  $|v_n|>v_c^T$ always holds and }
\be 
\label{expr-dvpos-snpos}
\Delta v_n=\frac{\sqrt{c_T^2-4J_n^2}}{|h_n|}>0\,,~~~s_n=-2+\frac{2+c_T}{h_n^2}=\frac{1+c_T+J_n^2}{h_n^2}>0\,.
\ee 
This corresponds to a step-wise increasing $|v_n|$ that we have not found either.

\begin{figure}[htbp]
    \centering
    \includegraphics[width=0.468\textwidth]{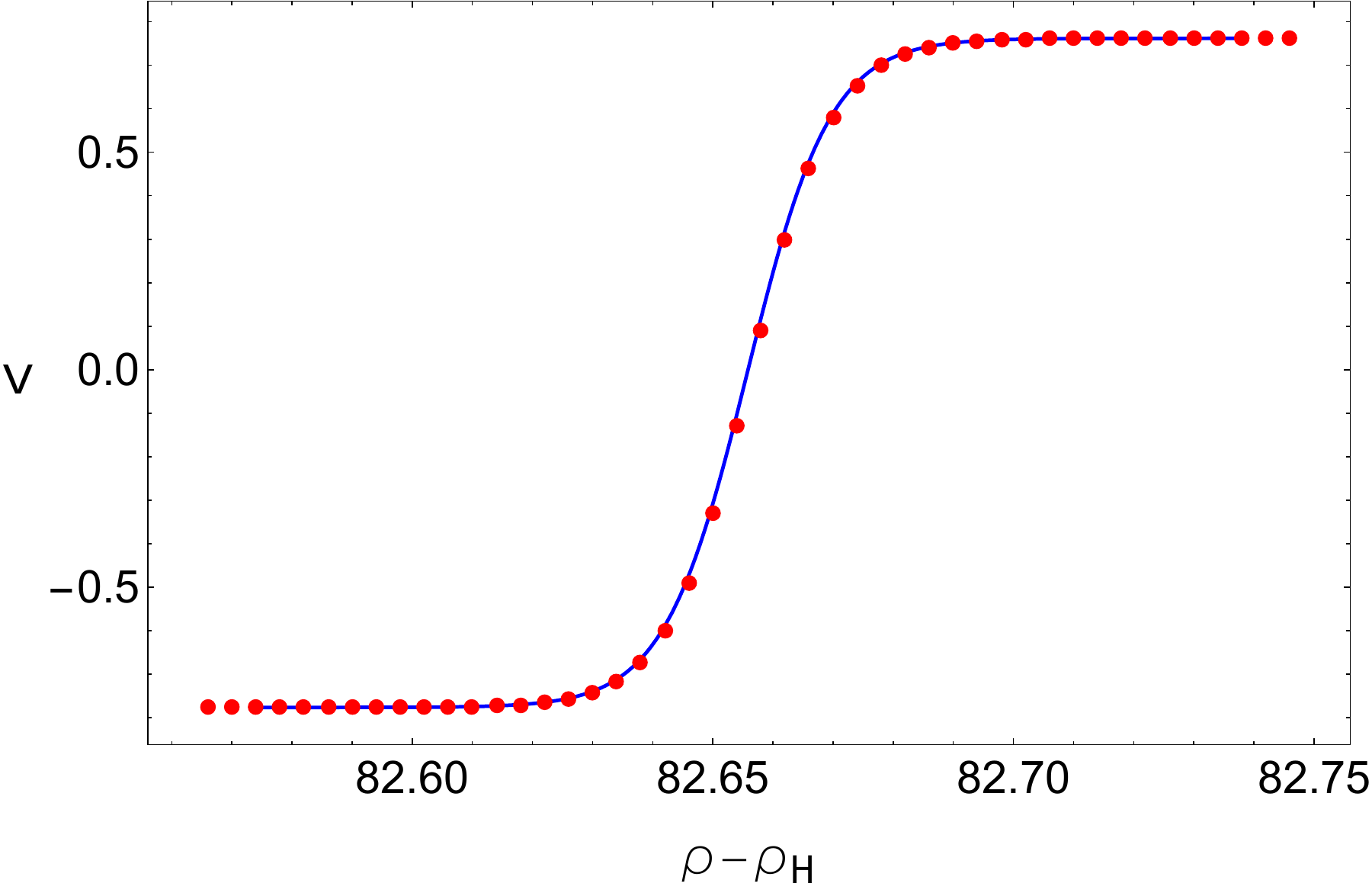}~~~
    \includegraphics[width=0.45\textwidth]{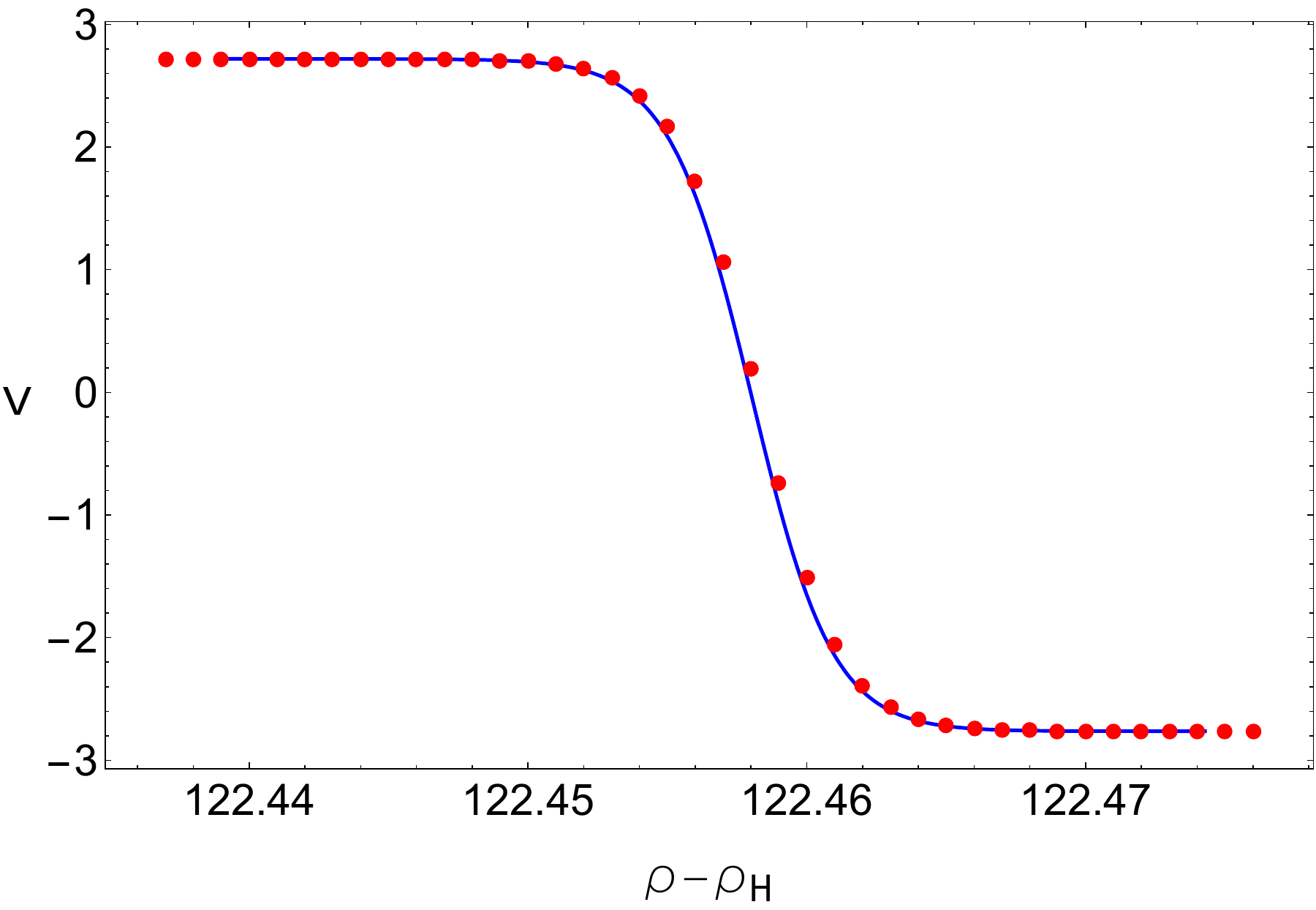}
    \vspace{-0.3cm}
    \caption{\small Evolution of $v$ as a function of $\rho$ within two types of Kasner transition. A bounce refers to the curve between two plateaus. The left figure shows a decreasing transition from $v_n=-0.777$ to $v_{n+1}=0.762$, while the right shows a increasing one from $v_n=2.719$ to $v_{n+1}=-2.763$. Solid blue lines represent the analytical results of \eqref{expr-v-T} and \eqref{expr-T}. We have $\rho_n=82.656\,,h_n=-0.0106 \,,c_T=-1.99984\,$ for the left figure and $\rho_n=122.458\,,h_n=0.008 \,,c_T=-2.00035\,$ for the right. The red points are the numerical results.}
    \label{fig-two-types}
\end{figure}

A brief summary of the dynamics of $|v|$ is as follows:  whether $|v|$ increases or decreases 
between two neighboring epochs 
depends on the relative magnitude of its initial value (i.e. $|v_{in}|$), and the critical value (i.e. $v_c^T=\sqrt{2}$ in \eqref{eq:criticalv}). %that is exactly determined by the Kasner transitions.
If we have $|v_{in}|<v_c^T$ for one transition, then $\left|v_{n+1}\right|<\left|v_{n}\right|\,$ holds for any value of $n$. While if $|v_{in}|>v_c^T$
then we have $\left|v_{n+1}\right|>\left|v_{n}\right|\,$ for any value of $n$. As a result, without dropping any term from the equation \eqref{eq-v-simp} for the field velocity $v$, we find two types of transitions, one is increasing and the other is decreasing. These are consistent with our numerical results, as shown in  Fig. \ref{fig-two-types}.

 {Note that the above discussion regarding whether $v$ between two  neighboring epochs increases or decreases, is only valid for Kasner transitions. Since a Kasner inversion may abruptly raise $|v_n|$ at some $n$ by \eqref{expr-v-inver}, the full interior evolution is determined by the co-effect of inversion and transitions, more precisely, by the critical values of $v_c^I$ and $v_c^T$. It is remarkable that the two critical values are equal,\footnote{ The critical value $v_c^I$ is independent of the choice of parameters in potential \eqref{expr-potential-generic}, since all potential terms can be neglected in the vicinity of inversion. $v_c^T$ is also independent of the choice, since the analytical derivation of $v_c^T$ remains the same under the reasonable assumptions \eqref{eq:Hproperity} with \eqref{expr-assump-hn}.} which leads to the full interior evolution not only simple but also highly interesting.} If the initial value $\abs {v_{in}} <v_c^T$, then the interior undergoes a series of Kasner transitions, during which $\abs v$ becomes progressively smaller than $v_c^T$. %Up to a 
Once $\abs v$ reaches a certain value 
$\abs {v_0}$, the rotation terms become non-negligible as $\rho$ increases, leading to a Kasner inversion. As a result, $\abs v$ is raised abruptly above $v_c^I$. The interior then undergoes infinitely many transitions, with $\abs v$ increasing monotonically as illustrated in Fig. \ref{fig-inv-tran}. {Since $\abs v$ is always greater than $v_c^I$ thereafter, no further inversion occurs as discussed in Sec. \ref{sec-inv}.} Similarly, if the initial value $\abs {v_{in}}>v_c^T$, then $\abs v$ becomes progressively larger than $v_c^T$. In this case, no inversion occurs, and the interior undergoes only infinitely many transitions, which is described in Fig. \ref{fig-no-inv-tran}.

It is important to emphasize that there are  at most two types of Kasner transitions in $|v|$ between two  neighboring epochs: monotonically increasing and decreasing. This is because the critical value for the inversion exactly coincides with that for the transitions, leading to the fact that the inversion triggered by rotation only alters the monotonous behavior of $\abs v$ before and after the inversion, without introducing any more complicated interior structures.
%The reason is that the critical value of the inversion is exactly the same as that of the transitions, under the reasonable assumption \eqref{expr-assump-hn} for a super-exponential potential.  
We do not focus extensively on the sign of $s_n$, which determines whether the scalar field oscillates or evolves  monotonically at late times.

\textcolor{black}{Now we make some remarks on the interior of 4D static black hole counterpart.} As discussed before, the inversion and transitions are independent processes. Therefore, within a single transition, we obtain the simplified equation \eqref{eq-v-simp} for $v$. This equation is the same as the one in 3D non-rotating black holes, inside which we expect two types of transitions depending on the initial value of $|v|$. Given the structural similarity between our equation \eqref{eq-v-simp} and that in \cite{Hartnoll:2022rdv}, we also expect these two types of transitions to exist in 4D counterpart. 
%Thus we also expect that the two types of transitions exist for the 4D counterpart in \cite{Hartnoll:2022rdv}, since our equation \eqref{eq-phi-tran} is very similar to theirs in structure.
The only difference lies in the coefficients. By repeating the same procedure in 4D, we find that all  expressions are highly similar to our results here. There the critical value of $v_c^T=2\sqrt{3}$ for transitions. Furthermore, we can predict the critical value of $v_c^I$ for inversion (if we assume the black hole is rotating or boosting along one spatial direction) by $v_c^I=v_c^T$, which has been confirmed in \cite{Prihadi:2025czn}. 

The reason why only decreasing type of transitions was found in \cite{Hartnoll:2022rdv} %they only find the decreasing type of transitions 
is that certain terms were omitted from \eqref{eq-v-simp}, resulting in 
a solution that only captures the decreasing transitions. {
Given that only decreasing transitions are found in 3D non-rotating black holes (Sec. \ref{sec: Comment on parameter spaces}), omitting these terms is plausible.
%Omitting these terms may not be problematic, as we also observe only decreasing transitions in 3D non-rotating black holes, as diccussed in Sec. \ref{sec: Comment on parameter spaces}.
}However,  for the increasing type of transitions (if they exist)  these terms should not be omitted. Moreover, in this case  the  singularity evolves towards a fundamentally different fate at extremely late times, as discussed in the following Sec. \ref{sec: Comment on late time singularity}. %This remains the case even though the parameter space supporting increasing transitions constitutes a set of very small measure within the full space of solutions.

\section{Analytical description of the late-time interior}
\label{sec:Late-time evolution}
\begin{figure}[htbp]
    \centering
    ~~
    \includegraphics[width=0.35\textwidth]{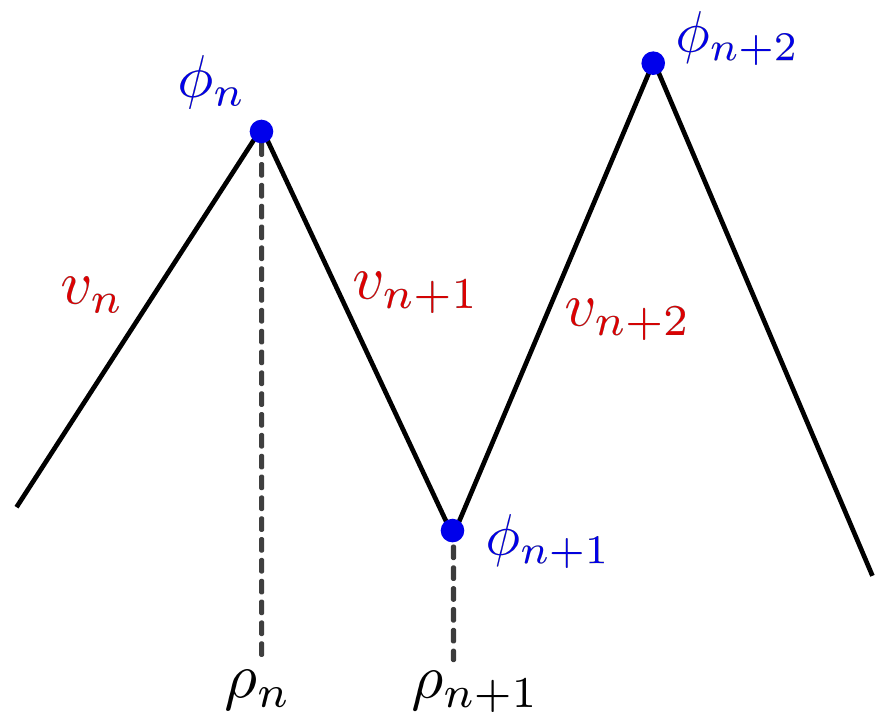}
    \hspace{35pt}
    %width=0.4\textwidth, ,height=0.34\textwidth
     \includegraphics[width=0.34\textwidth]{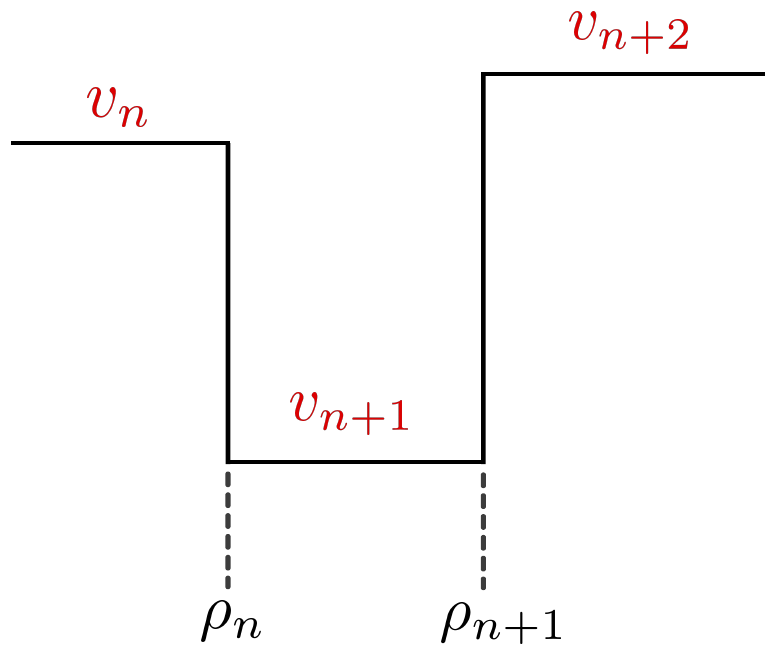}
    \vspace{-0.3cm}
    \caption{\small A schematic of the approximate evolution for $\phi(\rho)$ in $\phi$-$\rho$ plane ({\em left}) and $v$-$\rho$ plane ({\em right}). Note that $\phi(\rho)$ and $v(\rho)$ are related by  \eqref{trans-v}. In the left plot, $v_n$ is the slop of the $\phi(\rho)$ in the $n$-th epoch, while in the right plot, $\phi_{n}$ is the value of $\phi$ at $\rho_n$ and it satisfies \eqref{eq-rela1}. 
  The Kasner epochs range from $\rho_n$ to $\rho_{n+1}$, 
  allowing us to set up  recurrence relations for $\phi_n$, $v_n$ and $\rho_n$. 
  Assuming $\phi_n$, $v_n$ and $\rho_n$ have been obtained for arbitrary large $n$, then using the equation $\phi-\phi_n=v_{n+1}\left(\rho-\rho_n\right)$ we get the approximate $\phi(\rho)$ from $\rho_n$ to $\rho_{n+1}$. Thus we get the full evolution of $\phi(\rho)$. 
  The above analysis only applies when the widths of the bounces become progressively narrower than $\rho_{n+1}-\rho_n$, as we  demonstrate in \eqref{eq:width0}.
    }
    \label{fig-late-time-evolution-schematic}
\end{figure}

To understand the late-time evolution of the interior, it is essential to derive an  analytical solution for $\phi(\rho)$. Once $\phi(\rho)$ is derived, usually an approximate solution due to the extreme difficulty in obtaining an exact one, we can determine all metric functions using \eqref{eq-eom-tran}.\footnote{{Note that whereas $N$ is not constant during the inversion, 
it remains constant throughout the  transitions, as shown in Fig. \ref{fig-two-N}.}} Our strategy
proceeds as follows.  First, we establish three recurrence relations among 
the field velocities $v_n$, the field values at the bounces $\phi_n$, and the locations of the bounces $\rho_n$, where $n$ denotes the index number of the Kasner epoch (see Fig. \ref{fig-late-time-evolution-schematic}). 
By treating $n$ as a continuous variable at late times, we then solve for 
$v(n)$, $\phi(n)$ and $\rho(n)$. Therefore, once the initial value of $n$ is specified, we can determine $\phi(\rho)$ completely, and then evolve the entire spacetime. Fig.  \ref{fig-late-time-evolution-schematic} illustrates the procedure for the approximate evolution of $\phi(\rho)$.

Let us first derive three recurrence relations among $v_n$, $\phi_n$ and $\rho_n$. Since we only numerically find the case that $v$ and $\phi$ alternate in signs before and after a transition, we focus on the solutions where $\phi_n \phi_{n+1}<0$, and $v_n v_{n+1}<0$, i.e., \eqref{expr-vn-ct-neg} with \eqref{expr-dvpos-snneg} and \eqref{expr-dvneg-snneg}.
Thus the first relation 
\be 
\label{eq-rela1}
\phi_{n+1}-\phi_n=v_{n+1}\left(\rho_{n+1}-\rho_n\right)\,,
\ee 
is simplified to
\be 
\label{eq-rela1-abs}
\abs{\phi_{n+1}}+\abs{\phi_n}=\abs{v_{n+1}}\left(\rho_{n+1}-\rho_n\right)\,.
\ee 
The second relation for $v_n$ follows from \eqref{expr-vn-ct-neg},
\be 
\label{eq-rela2}
4\left(\abs{ v_{n+1} }-\abs{ v_n }\right)+8\abs{h_n}+\abs{h_n} \left(\abs{ v_{n+1} }-\abs{ v_n }\right)^2-\abs{h_n}
\left(\abs{ v_{n+1} }+\abs{v_n}\right)^2=0\,,
\ee
which is only valid for the solutions \eqref{expr-dvpos-snneg} and \eqref{expr-dvneg-snneg} we are considering. The derivation of the third relation is somewhat involved, as discussed below.
 
It is very useful to rewrite the equation \eqref{eq-phi-tran} via \eqref{trans-v} as 
\begin{equation}
			\begin{split}
			\label{eq-eomv-2}
		\frac{v''}{v'}=-\frac{2}{v}+\frac{H'-2v'}{H-2v}-\frac{v'}{v}+H-\left(v+\frac{2v'}{H-2v}\right)\,.
			\end{split}
\end{equation}
Note that primes denote derivatives with respect to $\phi$.
Integrating the above equation once we have
\be 
\label{eq-phidd}
2\rho+\log{\ddot{\phi}}-\log V_{sup}-\log{(H-2v)}+\int v\, \dd\phi+\int \frac{2v'}{H-2v} \,\dd\phi+c_l=0\,,
\ee 
where $c_l$ is a constant of integration.
The last two terms are not integrable, however, we note from Fig. \ref{fig-H} that $H$ could be regraded as a constant within a single bounce. Thus the last term could be approximately integrated to be
\be
\int \frac{2v'}{H-2v}\,\dd\phi \approx -\log{(H-2v)}\,.
\ee 
%though it is not entirely precise. 
Then \eqref{eq-phidd} can be rewritten as
\be 
\ddot{\phi}=c_l e^{-2\rho} V_{sup} (H-2v)^2 e^{-\int v^2\, \dd\rho}\,.
\ee 
Using the Laplace method we get 
\be 
\label{expr-vtot}
\begin{split}
v_{n+1}-v_n &=\int_{\rho_n^-}^{\rho_n^+}\ddot{\phi}\,\dd\rho=\pm c_l \int_{\rho_n^-}^{\rho_n^+} e^{-2\rho+\log{\left|V_{sup}(H-2v)^2\right|}-\int v^2 \,\dd\rho}\\ & \approx \pm \sqrt{2\pi \abs{c_l}}e^{-\rho_n}\sqrt{\abs{V_{sup}'(\phi_n)}} F(\rho_n)\,,
\end{split}
\ee 
where $F(\rho)$ is defined as 
\be 
F(\rho)\equiv e^{-\frac{1}{2}\int v^2 \,\dd\rho}\,,
\ee 
and we have used the following reasonable assumptions
\be 
\abs{\frac{H'}{H}}\ll \abs{H}\,,~~~~~\dot{\phi}(\rho_n)\approx 0\,.
\ee 
Note that we are considering the case $v_n v_{n+1}<0$, and thus we have
\be 
\left|v_{n+1}\right|+\left|v_{n}\right|=\sqrt{2\pi \abs{c_l}}e^{-\rho_n}\sqrt{\abs{V_{sup}'(\phi_n)}} F(\rho_n)\,.
\ee 
The difference $\rho_{n+1}-\rho_n$ is very useful,
\be 
\label{expr-rhodif-orig}
\begin{split}
\rho_{n+1}-\rho_n=&-\log{ \frac{\abs{v_{n+2}}+\abs{v_{n+1}}}{\abs{v_{n+1}}+\abs{v_n}}}+\log\sqrt{\abs{\frac{c_l^{n+1}}{c_l^n}}}\\
&~~+\log\sqrt{\abs{\frac{{V_{sup}'(\phi_{n+1})}}{{V_{sup}'(\phi_n)}}}}+\log F(\rho_{n+1})-\log F(\rho_{n})\,.
\end{split}
\ee 
It can be easily seen from the numerical results that the first two terms are far smaller than the other terms, and hence, \eqref{expr-rhodif-orig} is simplified to 
\be 
\label{expr-rhodif}
\rho_{n+1}-\rho_n=\log\sqrt{\abs{\frac{{V_{sup}'(\phi_{n+1})}}{{V_{sup}'(\phi_n)}}}}- \frac{1}{2}\int_{\rho_n}^{\rho_{n+1}} v^2 \,\dd\rho \,.
\ee 
The vecolity $v$ can be regarded as a constant $v\approx v_{n+1}$ from $\rho_n$ to $\rho_{n+1}$, as shown in Fig. \ref{fig-late-time-evolution-schematic}. Finally we get the third recurrence relation
\be
\label{eq-rela3}
\rho_{n+1}-\rho_n=\log\sqrt{\abs{\frac{{V_{sup}'(\phi_{n+1})}}{{V_{sup}'(\phi_n)}}}}- \frac{1}{2}(\rho_{n+1}-\rho_n)\abs{v_{n+1}}^2\,.
\ee 

\begin{figure}[h!]
    \centering
    \includegraphics[width=0.48\textwidth]{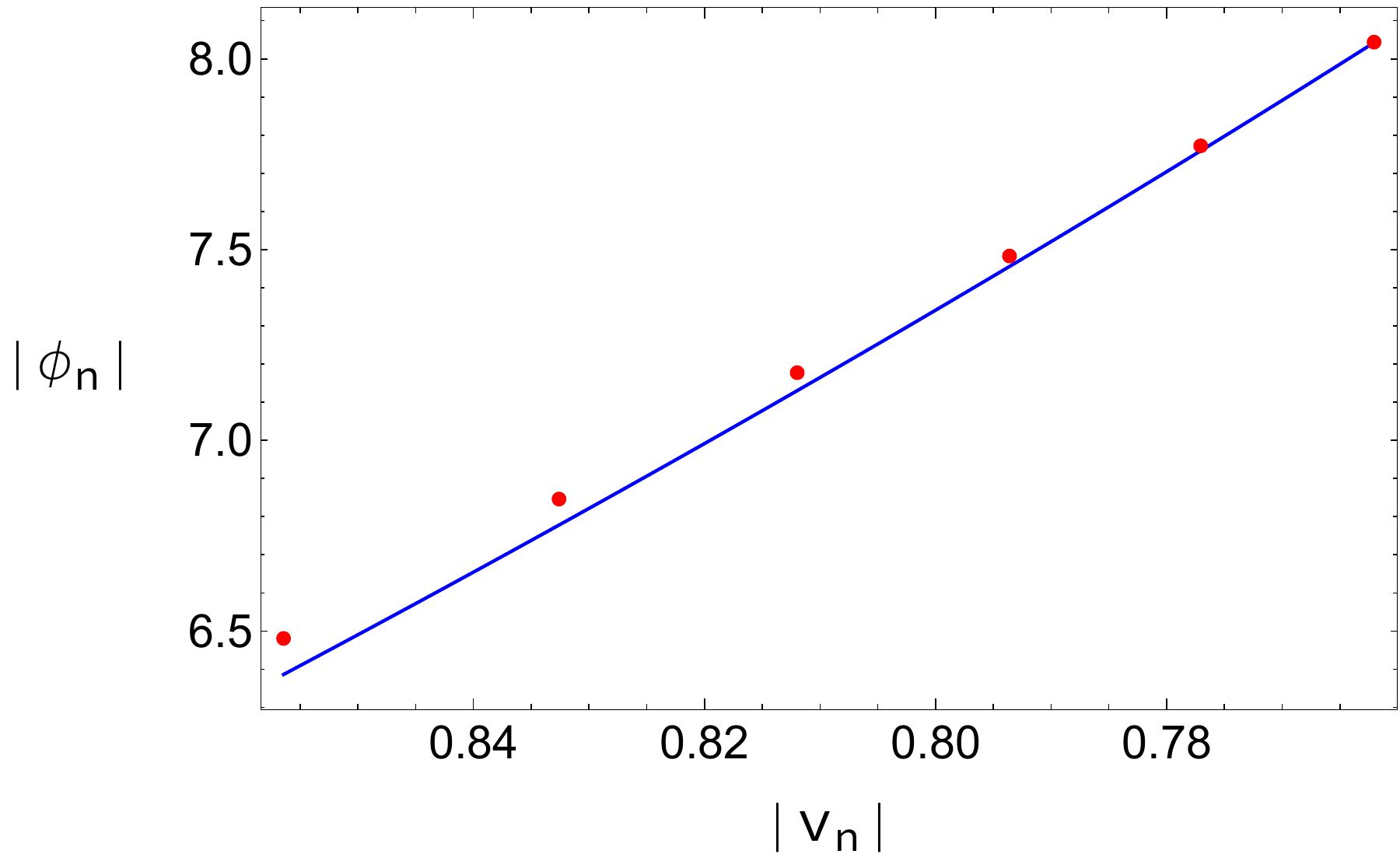}~~~
    \includegraphics[width=0.45\textwidth]{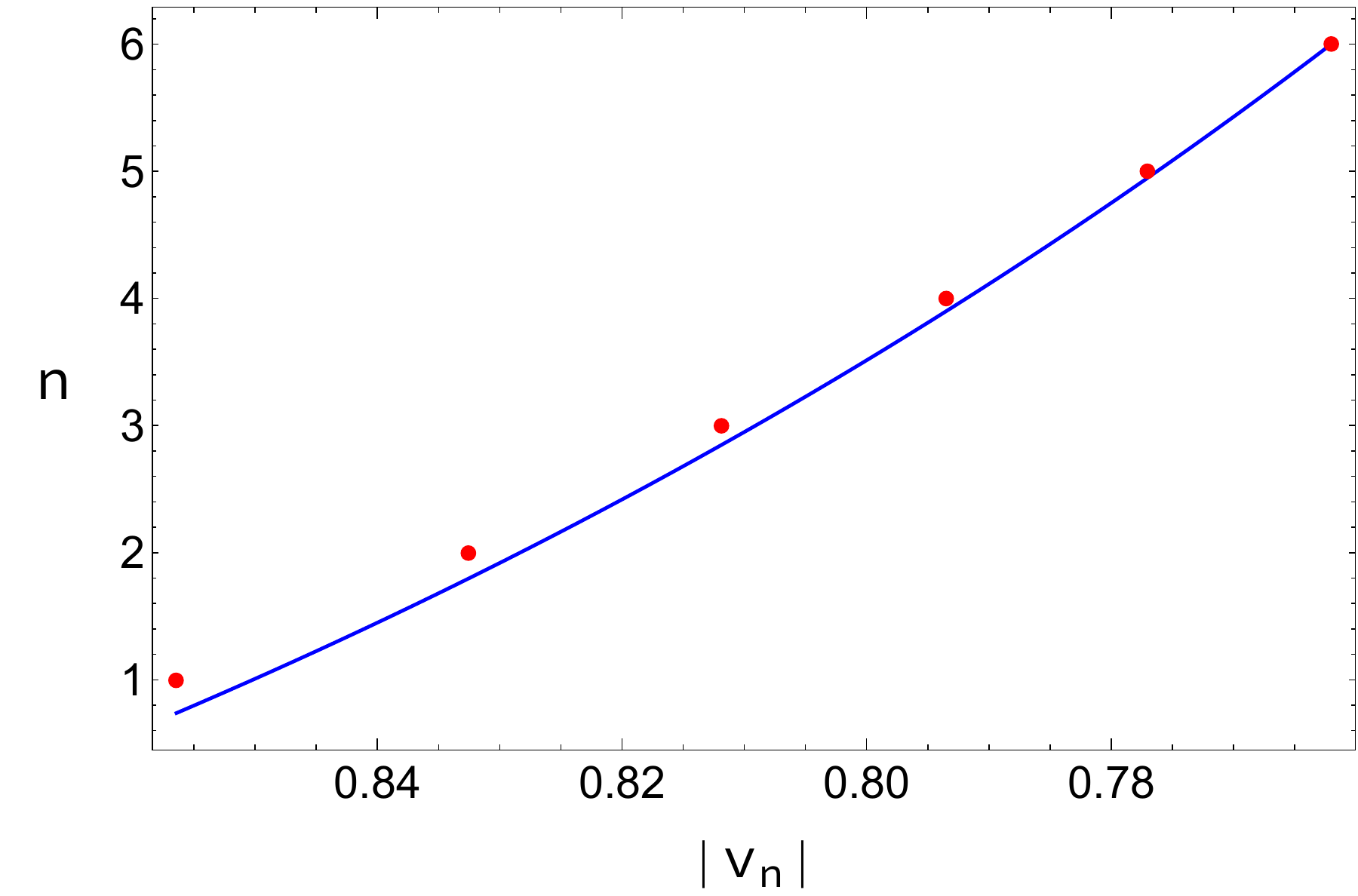}
    \vspace{-0.3cm}
    \caption{\small {Behavior of $|\phi_n|$ ({\em left}) and $n$ ({\em right}) as functions of $|v_n|$ for the decreasing transitions. The parameters  
    $b=4.318$, $n_0=-27.29$. Note that the $x$-axis is plotted in reverse (i.e., values decrease from left to right).}
    }
    \label{fig-match-phinv-no-inv}
\end{figure}

{At extremely late interior times $n\rightarrow \infty$, we may consider all functions are continuous with respect to $n$, i.e. $\abs{v_n}=v(n)$, $\abs{\phi_n}=\phi(n)$, and $\abs{\rho_n}=\rho(n)$.\footnote{It turns out the analytical solution also works well at small $n$, as we will show in Fig. \ref{fig-match-phinv-no-inv}.} We can safely make the following approximations}
\be
{
\abs{v(n+1)-v(n)}\ll \frac{4}{\abs{h_n}}\,,~~\abs{\frac{v(n+1)-v(n)}{v(n)}}\ll 1\,,~~\abs{\frac{\phi(n+1)-\phi(n)}{\phi(n)}}\ll 1\,,}
\ee 
{which are well satisfied with numerical data.
Thus the three recurrence relations 
\eqref{eq-rela1-abs},\eqref{eq-rela2} and \eqref{eq-rela3}
are simplified to}
\be 
\label{expr-threeRelations-simp}
\begin{split}
&\frac{2\phi(n)}{v(n)}=\rho(n+1)-\rho(n)\,,\\&v(n+1)-v(n)=\abs{h(n)}\left(v(n)^2-2\right)\,,\\ &\rho(n+1)-\rho(n)=\log\sqrt{\abs{\frac{{V_{sup}'(n+1)}}{{V_{sup}'(n)}}}}- \frac{1}{2}\left(\rho(n+1)-\rho(n)\right)v(n)^2\,.
\end{split}
\ee 
We can use the first equation to eliminate $\rho(n)$ and get the two equations for $\phi(n)$ and $v(n)$,
\be 
\begin{split}
&v(n+1)-v(n)=\abs{h(n)}\left(v(n)^2-2\right)\,,\\ &\frac{2\phi(n)}{v(n)}=\log\sqrt{\abs{\frac{{V_{sup}'(n+1)}}{{V_{sup}'(n)}}}}-\phi(n) v(n)\,,
\end{split}
\ee 
which can be further transformed into two differential equations,
\be 
\label{eq-diff}
\begin{split}
\frac{\mathrm{d} v}{\mathrm{d}n}&=\abs{h(n)}(v^2-2)\,,\\ \frac{\mathrm{d}\phi}{\mathrm{d}n}&=\abs{h(n)}\left(\frac{2\phi}{v}+\phi v\right)\,.
\end{split}
\ee 
We immediately get
\be 
 \phi= \frac{b \left(2-v^2\right)}{v}\,,
    \label{expr-phiv}\\
\ee 
where $b$ is a constant of integration that depends on the types of transitions.\footnote{Note that after \eqref{eq-diff} until \eqref{expr-rhov} in this section, $\phi$, $v$ and $\rho$ are functions of $n$ rather than those of $z$ or $\rho$.} For given values of $\phi_n$ and $v_n$, the  parameter $b$ can be determined, which yields an expression valid for arbitrary large $n$. %\footnote{It turns out that at smaller value of $n$, the above approximation also works, as shown in Fig.  \ref{fig-match-phinv-no-inv}.}
\iffalse
In the limit of $v\rightarrow 0$\,,
\be 
\phi=\frac{2b}{v}\,
\ee 
with $b>0$, which corresponds to the decreasing transitions. 
And in the limit of $v\rightarrow \infty$\,, 
\be 
\phi=-b v\,
\ee 
with $b<0$, which corresponds to the increasing transitions. 
\fi

{Note that we have parameterized the super-exponential potential \eqref{expr-potential-generic} to be $V_{sup}=\, e^{\int H \,\dd\phi}$ in \eqref{expr-potential-repara}, which corresponds to the power-law function $H=\lambda \phi^l$ with $\lambda=2k_3k_4$ and $l=2k_4-1$. The generic solutions to \eqref{eq-diff} are}
\begin{align}
    n&=n_0+\frac{\lambda b^l (2-v^2)^l}{4 v^l (l-1)}\, {}_2F_1\left(1,\frac{1+l}{2};\frac{3-l}{2};\frac{v^2}{2}\right)\,. \label{expr-nv}
\end{align}
In order that the scalar field cannot escape to infinity in either direction, we have restricted to even potentials, i.e., $l$ is odd.\footnote{For odd super-exponential  potentials, $v$ may evolves into a constant, since the super-exponential potential could be significantly suppressed by negative $\phi$.} For example, we have 
\iffalse
When $l=2s$ and $s \in \mathbb{N}^+$, \eqref{expr-nv} is in the form of a power series, such as
\be 
\label{expr-leven}
\begin{split}
    n-n_0&=\frac{\lambda b^2}{2} \frac{\left(2+v^2\right)}{v}\,,~~l=2\,.\\
    n-n_0&= \frac{\lambda b^4}{6}\frac{\left(8-36v^2-18v^4+v^6\right)}{v^3}\,,~~l=4\,.\\
     n-n_0&= \frac{\lambda b^6}{30}\frac{\left(96-400v^2+1200v^4+600v^6-50v^8+3v^{10}\right)}{v^5}\,,~~l=6\,.
\end{split}
\ee 
\fi
\be 
\label{expr-nv-lodd}
\begin{split}
    %n-n_0&=-\frac{1}{2}\lambda b \log v\,,~~l=1\,.\\
    n-n_0&=-\frac{\lambda b^3}{4} \frac{\left(-4+v^4\right)}{v^2}+2 \lambda b^3 \log v\,,~~l=3\,.\\
     n-n_0&=-\frac{\lambda b^5}{8} \frac{\left(-16+64v^2-16v^6+v^8\right)}{v^4}- 12 \lambda b^5 \log v\,,~~l=5\,.\\
\end{split}
\ee 
It can be easily seen that as $n\rightarrow \infty$, we have two different limits for $v$, $v\rightarrow 0$ that corresponds to the decreasing transitions and $v\rightarrow \infty$ for the increasing transitions. More precisely,\footnote{We do not discuss the case $0<l<1$, since it 
corresponds to the case that $\phi(n)$ is finite as $n\rightarrow \infty$, which could destroy the assumption that $\abs{H(\phi_n)}\gg 1$. }
\be 
\label{expr-nv-v0}
 n-n_0=\frac{2^{-2+l}}{l-1} \frac{\lambda b^l}{v^{l-1}}\,,~~l\in \mathbb{N}^+\,,l>1\,,~~\text{as}\quad v\rightarrow 0\,,
\ee
and
\be 
\label{expr-nv-vinfty}
 n-n_0=\frac{-\lambda b^l}{2(l-1)} v^{l-1}\,,~~l\in \mathbb{N}^+\,,l>1\,,~~\text{as}\quad v\rightarrow \infty\,.
\ee

{ The expressions from \eqref{expr-phiv} to  \eqref{expr-nv-vinfty} are valid for a broad class of potentials \eqref{expr-potential-generic}. We can derive $\rho(v)$ by plugging \eqref{expr-phiv} and \eqref{expr-nv-lodd}
into the first equation in \eqref{expr-threeRelations-simp}, and solve the corresponding differential equation. 
For the potential  \eqref{expr-potential-generic} with $l=3$, we have %For $l=3$ we have
\be 
\label{expr-rhov}
\rho-\rho_0=b^4\lambda\left(\frac{4-12v^2+v^6}{2v^4}-6\log v\right)\,.
\ee
Together with the expressions \eqref{expr-phiv} and \eqref{expr-nv-lodd}, we have the full analytical solution for the interior of black hole in theory with the specific potential \eqref{expr-potential}. 
}

\begin{figure}[h!]
    \centering
    \includegraphics[width=0.468\textwidth]{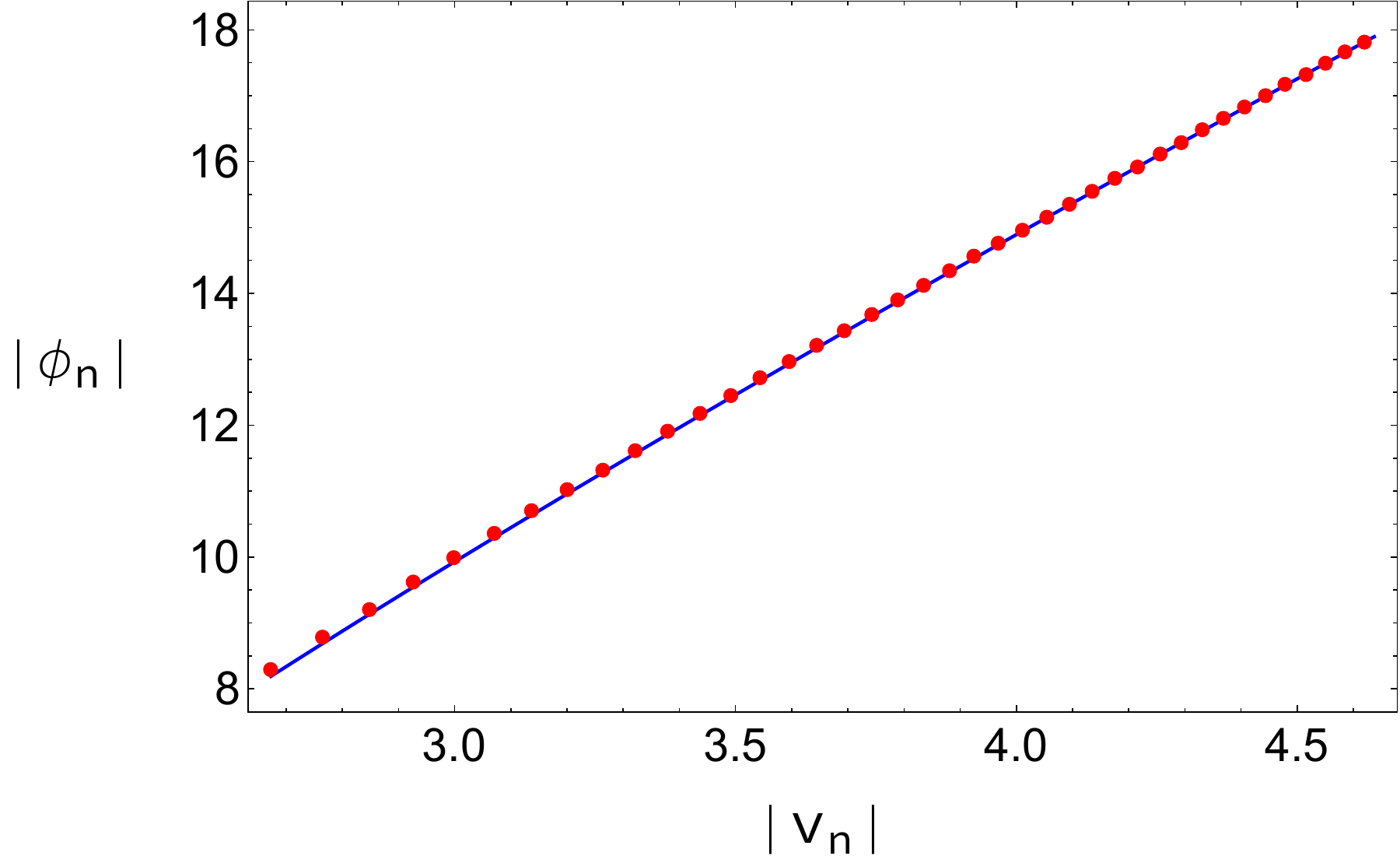}~~~
    \includegraphics[width=0.45\textwidth]{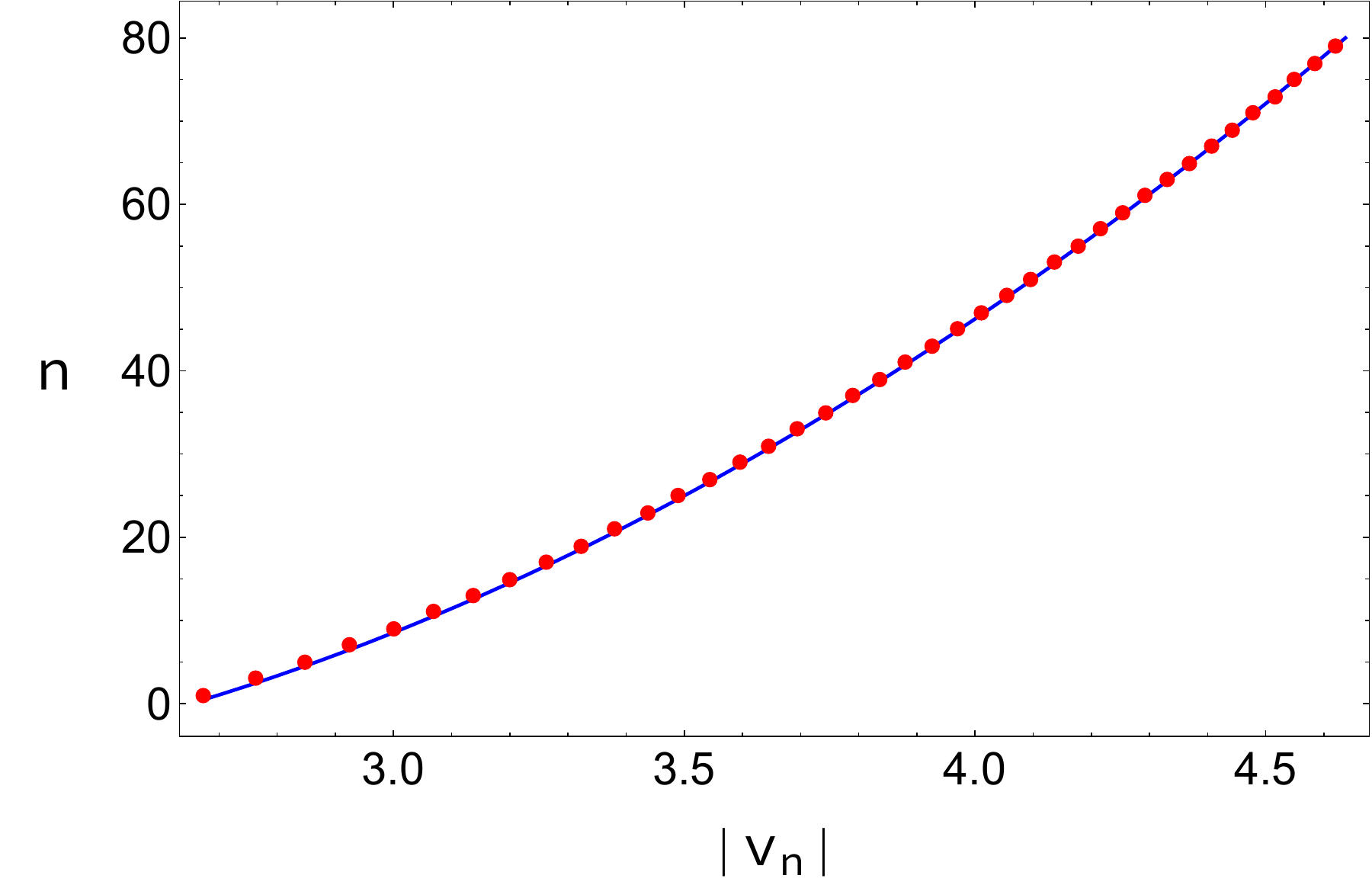}
    \vspace{-0.3cm}
    \caption{\small Behavior of $|\phi_n|$ ({\em left}) and $n$ ({\em right}) as functions of $|v_n|$ for the increasing transitions. The red dots are numerical results, while the blue lines are analytical curves from  \eqref{expr-phiv} and  \eqref{expr-nv-lodd} with the parameters  
    $b=-4.255$, $n_0=10.34$. }
    \label{fig-match-phiv-&-vn}
\end{figure}

{With the specific potential \eqref{expr-potential}, we can numerically verify our analytical results. Fig. \ref{fig-match-phiv-&-vn} and Fig. \ref{fig-match-phinv-no-inv} show the exact agreement between the analytical expressions \eqref{expr-phiv}, \eqref{expr-nv-lodd} and the numerical results. The solution \eqref{expr-rhov} is also consistent with the numerical data, which we do not show here for brevity.}

Additionally, as $n$ increases, the effective widths of bounces derived from \eqref{expr-vappr} go to zero,\footnote{{It follows from Fig. \ref{fig-two-types} that there is a curve connecting the left Kasner epoch and the right one. The curve refers to the bounce, and the width of the bounce refers to the width of the curve.}}
\be 
\label{eq:width0}
\left|\frac{(-1+J_n^2)\sqrt{c_2^2+4J_n^2}}{2c_T J_n^2}\right|=\frac{|h_n|\sqrt{4+(-8+{\delta v}_n^2)h_n^2}}{{\delta v}_n(1-2h_n^2)}\approx \frac{2|h_n|}{{\delta v}_n}=\frac{4}{|H_n| {\delta v}_n}\approx \frac{2}{|H_n v_n|}\rightarrow 0\,
\ee 
with $\delta v_n=|v_{n+1}-v_n|=|v_{n+1}|+|v_n|$, while the widths of Kasner epochs \(\left|\rho_{n+1}-\rho_n\right|\) derived from  \eqref{expr-nv-lodd} and \eqref{expr-rhov} can be easily proved to be finite or divergent. {Therefore, deep in the interior, these bounce regions can be safely neglected. Then using the relation 
\be \phi-\phi_n=v_{n+1}\left(\rho-\rho_n\right)\,,
\ee 
we can get the approximate $\phi(\rho)$ within a single Kasner epoch from $\rho_n$ to $\rho_{n+1}$. Since the  expressions \eqref{expr-phiv}, \eqref{expr-nv}, \eqref{expr-rhov} allow us to derive $\phi_n$, $v_n$ and $\rho_n$ for all epochs, we can get the full evolution of $\phi(\rho)$ at late times, and then evolve the entire spacetime.}
%Since we have derived the expressions for  $\phi_n\,, v_n$ and $\rho_n$, as shown in \eqref{expr-phiv}, \eqref{expr-nv}, \eqref{expr-rhov}, we can get the full evolution of $\phi(\rho)$ at late times.} 

\vspace{0.3cm}
To determine the nature of the singularity, we also need to derive the expressions for $f_K^n $ and $\chi_1^n$ in \eqref{expr-Kas-fchiN}. %for use in the following subsection. 
Note that we have obtained the approximate solution \eqref{expr-vappr} for $v(\rho)$, plugging which into the simplified equations of motion \eqref{eq-eom-tran} within a single transition we get the corresponding equations of motion for $\chi(\rho)$ and $f(\rho)$,
\be 
\begin{split}
\label{eq:chi-f-rho}
\dot{\chi}&=2  (v^T_{appr})^2\,,\\[10pt]
\frac{\dot{f}}{f}&=\frac{2-2h_n v^T_{appr}+(v^T_{appr})^2-h_n (v^T_{appr})^3+h_n\dot{v}^T_{appr}}{1-h_n v^T_{appr}}\,,
\end{split}
\ee 
where dots denote derivatives with respect to $\rho$, namely the coordinate itself, not the function $\rho(n)$. The full solutions are lengthy and the detailed formulae can be found in Appendix \ref{app:sol}. Defining 
\be 
f_K(\rho)\equiv-\frac{f(\rho)}{z^{2+(v^T_{appr})^2}}\,,~~~\chi_1(\rho)\equiv\chi(\rho)-2(v^T_{appr})^2 \log z\,,
\ee 
within a single transition we have 
\be
\label{eq:chinfkn}
\begin{split}
\chi_1^n=\lim_{\rho \to -\infty}\chi_1(\rho)\,,~~~\chi_1^{n+1}=\lim_{\rho \to +\infty}\chi_1(\rho)\,,\\[4pt]
f_K^n=\lim_{\rho \to -\infty}f_K(\rho)\,,~~~f_K^{n+1}=\lim_{\rho \to +\infty}f_K(\rho)\,.
\end{split}
\ee
Thus we have
\be\label{eq:relation-chinfn} 
\begin{split}
\chi_1^{n+1}-\chi_1^n&=\frac{4 \left(c_T+2\right) \rho _n \sqrt{c_T^2-4 J_n^2}}{J_n^2-1}-\frac{c_T \sqrt{c_T^2-4 J_n^2}}{J_n^2}\,,\\
\log{\frac{f_K^{n+1}}{f_K^n}}&=\frac{1}{2}(\chi_1^{n+1}-\chi_1^n)-2 \tanh ^{-1}\left[-\frac{\sqrt{c_T^2-4J_n^2}}{c_T}\,\right]\,.
\end{split}
\ee 
Using the expression for $v(n)$ in \eqref{expr-vn-ct-neg} to eliminate $c_T$ and expressing $J_n$, $h_n$ and $\rho_n$ in terms of $v(n)$, we can get $\chi_1^{n+1}-\chi_1^n\,$ and $\log f_K^{n+1}-\log f_K^n\,$ as functions of $v(n)$. The full expressions are unnecessarily cumbersome to present here. 

We shall focus on the behavior of  $f_K^n $ and $\chi_1^n$ in the limit $n\to\infty$, which corresponds to $v(n)\rightarrow\infty$ or $v(n)\rightarrow 0$. 
For the case $l=3$, in the limit  $v(n)\rightarrow\infty$ we have
\be 
\begin{split}
\chi_1^{n+1}-\chi_1^n&=4 b  v(n)^2-48b\log{v(n)}+\left(16b-\frac{12}{b^2\lambda}+\frac{8\rho_0}{b^3\lambda}\right)+...\,,\\
\log{\frac{f_K^{n+1}}{f_K^n}}&=2 b  v(n)^2-24b\log{v(n)}+\left(8b-\frac{6}{b^2\lambda}+\frac{4\rho_0}{b^3\lambda}\right)+...\,.
\end{split}
\ee 
While in the limit $v(n)\rightarrow 0$,
\be 
\begin{split}
\chi_1^{n+1}-\chi_1^n&=4b+v(n)^2
\left(-8b-\frac{1}{b^2\lambda}\right)-12b\,v(n)^4\log{v(n)}+...\,,\\
\log{\frac{f_K^{n+1}}{f_K^n}}&=2b+v(n)^2\left(-4b-\frac{1}{2b^2\lambda}\right)-6b\,v(n)^4\log{v(n)}+...\,.
\end{split}
\ee 
Thus we can get the approximate differential equations for $\chi_1(n)$ and $f_K(n)$ by replacing $\chi_1^{n+1}-\chi_1^n$ with $\mathrm{d}\chi_1/\mathrm{d}n$, and $\log f_K^{n+1}-\log f_K^n$ with $\mathrm{d}(\log f_K)/\mathrm{d}n$.
In the limit $v(n)\to \infty$, the leading behaviors of the solutions are
\be 
\begin{split}
\label{expr-chif-vinfty}
\chi_1=&\,c_1-\frac{1}{2} b^4 \lambda  v^4+v^2 \left(-6 b^4 \lambda +12 b^4 \lambda  \log v+3 b-2 \rho _0\right)-36 b \log ^2 v\\&~~~+4 \log v \left(6 b^4 \lambda +\frac{3 \rho _0}{b^3 \lambda }-\frac{4}{b^2 \lambda }+8 b-1\right)\,,\\
\log f_K=&\,c_2-\frac{1}{4} b^4 \lambda  v^4+v^2 \left(-3 b^4 \lambda +6 b^4 \lambda  \log v+\frac{3 b}{2}-\rho _0\right)-18 b \log ^2 v\\&~~~+2\log v \left(6 b^4 \lambda +\frac{3 \rho _0}{b^3 \lambda }-\frac{4}{b^2 \lambda }+8 b-2\right)\,,
\end{split}
\ee 
where $c_1$ and $c_2$ are two constants of integration. While in the limit $v(n)\to 0$,\footnote{Since we do not have sufficient data for decreasing transitions in the interior shown in Fig. \ref{fig-inv-tran}, the subleading terms $v^2$ and $v^4$ are essential for ensuring the high accuracy of the solutions, although they will not have an effect on determining the nature of the singularity in the next subsection.} 
\be 
\label{expr-chif-v0}
\begin{split}
   \chi_1=&\, c_3+\frac{4 b^4 \lambda }{v^2}+2 \left(12 b^4 \lambda +b\right) \log v
   +v^2 \left(-6 b^4 \lambda +12 b^4 \lambda  \log v-3 b-2 \rho _0+1\right)\\&~~~+v^4 \left(\frac{-4 b^7 \lambda ^2+2 b^3 \lambda +2 b+2 \rho _0-1}{8 b^3 \lambda }-\frac{3}{2} b \log v\right)\,,\\
   \log f_K=&\,c_4+\frac{2 b^4 \lambda }{v^2}+\left(12 b^4 \lambda +b\right) \log v+v^2 \left(-3 b^4 \lambda +6 b^4 \lambda  \log v-\frac{3 b}{2}-\rho _0+1\right)\\
   &~~~+v^4 \left(\frac{-2 b^7 \lambda ^2+2 b^3 \lambda +b+\rho _0-1}{8 b^3 \lambda }-\frac{3}{4} b \log v\right)\,,
\end{split}
\ee 
where $c_3$ and $c_4$ are two constants of integration.
We have verified that the above analytical solutions agree well with the numerical results. These solutions are useful for determining  whether the interior evolves towards a curvature singularity at extremely late times.
%xxxxxxxxxxxxxxxxxxxxxxxxxxxxxxxxxxxxxxxxxxxxx

\subsection{Behavior of the late-time singularity}
\label{sec: Comment on late time singularity}

The late-time evolution of the interior remains the same, regardless of whether an inversion occurs, as early shown in Fig. \ref{fig-inv-tran} and Fig. \ref{fig-no-inv-tran}. Hence, we restrict our attention to the case with inversion and study the interior after the inversion at late times. Note that we have verified that $N_K^f=N_h$ is always a constant after the inversion in Sec. \ref{sec-inv}. Performing the transformation
\be 
\label{expr-tran-xtoy}
y=N_K^f t+x\,,
\ee 
where $y$ has the same period as $x$, namely $2\pi$,
within a single Kasner epoch the metric \eqref{expr-ds-kas} reads 
\be 
\label{expr-ds-kas-noN}
\dd s^2=-\dd\tau^2+c_t \tau^{2p_t}\dd t^2+ c_x \tau^{2p_x} \dd y^2\,.
\ee 
During the transitions from one Kasner epoch to the next, the field velocity $v$ approaches infinity at extremely late-times, as indicated in \eqref{expr-nv-vinfty}.  Thus, it follows from \eqref{expr-kas-exponents} that after the inversion, the Kasner exponents evolves towards
\be 
\label{expr-kas-vinfty}
p_t\rightarrow 1\,,~~~p_x\rightarrow 0\,,~~~p_{\phi}\rightarrow 0\,.
\ee  

Firstly, we aim to examine the spacetime described by the Kasner metric \eqref{expr-ds-kas-noN} with exactly \(p_t=1\) and \(p_x = 0\). It is not the singularity in non-rotating BTZ black holes with \(p_t=0\) and \(p_x=1\), despite having the same local geometry. The metric can be deformed into
\be 
\label{expr-milne-metric}
\dd s^2=-\dd\tau^2+\tau^2 \dd\tilde{t}\,^2+ c_x \dd y^2\,,
\ee 
which is the two dimensional Milne universe cross a circle, namely, \(\text{Milne}_{1+1} \times S^1\,.\) There is no curvature singularity since all curvature invariants are finite. We can also prove that no causal singularity, which appears in non-rotating BTZ black holes, arises. Assuming a closed timelike curve $\tau(\lambda)\,,\tilde{t}(\lambda)\,,y(\lambda)$ exists, we have 
\be 
\dd s^2=-\frac{\dd\tau^2}{\dd{\lambda}^2}+\tau^2 \frac{\dd\tilde{t}\,^2}{\dd{\lambda}^2}+ c_x \frac{\dd y^2}{\dd{\lambda}^2}<0\,.
\ee 
Since $\tau$ will come back to its value at $t_0$, it has to be such that $\dd\tau/\dd\lambda=0$ at some value of $\lambda$. Then at this point, $\dd\tilde{t}/\dd\lambda$ and $\dd y/\dd\lambda$ must also be zero so that the curve terminates at this point, and a contradiction arises. Actually, $\text{Milne}_{1+1}$ universe only covers half of $1+1$ dimensional Minkowski spacetime. %$\text{Minkowski}_{1+1}$.   
Even if we extend the $\text{Milne}_{1+1}\times S^1$ to the other patch of $\text{Minkowski}_{1+1}\times S^1$, which corresponds to the region $\tau^2<0$, there is still no closed timelike curve. 

One might naively expect that this would provide another approach to realize black holes with regular interiors beyond the setup in 
\cite{Bueno:2025dqk}.
However, the following careful examination of the curvature of \eqref{expr-ds-kas-noN} with \eqref{expr-kas-vinfty} shows it actually diverges as $\tau\rightarrow 0$. It turns out that the key mechanism determining the nature of the singularity is the non-commutativity between the \(\tau \to 0\) and \(p_t \to 1\) (\( v \to \infty\)) limits.

The curvature of \eqref{expr-ds-kas-noN} reads 
\be 
R=R_{\mu\nu} R^{\mu\nu}= \frac{2 p_t(p_t-1)}{\tau^2}\,,~~~R_2=R_{\mu\nu\alpha\beta} R^{\mu\nu\alpha\beta}=3R^2\,.
\ee
It is sufficient to focus only on the late-time behavior of $R$, which can be expressed in terms of $v$,
\be 
R(\tau;v)=-\frac{4v^2}{(2+v^2)^2 \tau^2}\,.
\ee 
If we first take the limit $v\rightarrow \infty$, followed by $\tau \rightarrow 0$, then it indeed describes a singularity evolving towards the regular spacetime $\text{Milne}_{1+1}\times S^1$. But our case is far different. Within a given Kasner regime, the proper time $\tau$ is not free to approach to zero. Or equivalently the coordinate $z$ is not free to approach infinity. Instead, it is constrained to lie in a range approximately from $e^{\rho_{n-1}}$ to $e^{\rho_n}$, which is supported by \eqref{eq:width0} that the effective widths of bounces vanish. Thus $\tau$ is constrained by \eqref{tran-tauToz} in the range $(\tau_{n-1},~\tau_n)$
with
\be 
\label{expr-taun}
\tau_{n-1}=\frac{2}{\sqrt{f_K^n}(v_n^2+2)}e^{-\rho_{n-1}(v_n^2+2)/2}\,,~~~ \tau_n=\frac{2}{\sqrt{f_K^n}(v_n^2+2)}e^{-\rho_n(v_n^2+2)/2}\,.
\ee
Note that $|R|$ increases monotonically as $\tau$ decreases in the range. Thus we focus on the maximum of $|R|$,
\be 
R_m=|R(\tau_n,v_n)|=\frac{4v_n^2}{(2+v_n^2)^2\tau_n^2}\,.
\ee 
Plugging \eqref{expr-rhov}, \eqref{expr-chif-vinfty} and \eqref{expr-taun} into the above expression we have
\be
\begin{split}
\log {R_m}=&\frac{1}{4} b^4 \lambda  v^4+v^2 \left(\frac{3 b}{2}-2 b^4 \lambda \right)+\log v \left(\frac{6 \rho _0}{b^3 \lambda }-\frac{8}{b^2 \lambda }+16 b-2\right)
\\&~~~-18 b \log ^2 v+c_2-6 b^4 \lambda+2 \rho _0+O(\frac{1}{v^2})\,,
\end{split}
\ee 
which is again in excellent agreement with our numerical data. We can clearly see that as $v\rightarrow \infty$, the curvature increases progressively from one Kasner epoch to the next, and therefore the singularity remains a curvature one at late times.

Now we consider the case in which, before the inversion, the Kanser transition occurs large number of times.\footnote{We do not have to fine-tune the initial values at the horizon, whereas we only need to choose the initial values carefully.} Since the metric function $N$ is a constant, $N_K^i=1/N_h=-k/\omega$ as discussed in Sec. \ref{sec-inv}, within a single Kasner epoch the metric \eqref{expr-ds-kas} can also be deformed into \eqref{expr-ds-kas-noN}. We can see from \eqref{expr-nv-v0} that \(v \to 0\) at large $n$, and thus the Kasner exponents evolves towards
\be 
\label{expr-kas-v0}
p_t\rightarrow 0\,,~~~p_x\rightarrow 1\,,~~~p_{\phi}\rightarrow 0\,,
\ee  
which fortuitously exchanges $p_t$ and $p_x$ in \eqref{expr-kas-vinfty}. This is traceable. From \eqref{expr-v-inver}, we find that the inversion exchanges the values of $p_t$ and $p_x$ before and after the inversion. Here, similarly, the inversion exchanges the evolution directions of the transitions. The same question arises: whether the interior evolves towards the non-curvature singularity with \(p_t=0\) and \(p_x=1\), which is the singularity in non-rotating BTZ black holes. 
The analysis here is the same as that in the case $v\rightarrow \infty$. We have
\be 
\log {R_m}=\frac{4 b^4 \lambda }{v^4}-\frac{8 b^4 \lambda }{v^2}+(b+2) \log v+c_4-6 b^4 \lambda+2 \rho _0+O(v^2)\,,
\ee 
showing equally good agreement with our numerical results. The curvature becomes progressively larger as $v\rightarrow 0$, and the singularity still remains a curvature one. It should be emphasized that the above late-time behavior may appear only when the inversion exists. If the initial \(|v|>v_c^T\), no inversion occurs, and the interior only evolves towards \eqref{expr-kas-vinfty} while the curvature diverges.

\subsection{Comment on additional cases}
\label{sec: Comment on parameter spaces}

In the previous sections, we examined the co-effect of the rotation and a complex scalar field with a super-exponential potential \eqref{expr-potential}.
However, we find that in the presence of a super-exponential potential, the simplified equations of motion in the vicinity of a single transition are independent of whether the scalar field is real or complex, and also independent of whether the black hole is rotating or not. That means we can utilize \eqref{expr-phiv} and \eqref{expr-nv-lodd} to discuss the three additional cases: 

(i) $N$ is non-trivial and $\omega=k=0$: rotating black holes with a real scalar field,

(ii) $N=\omega=0$ and $k\neq0$: non-rotating black holes with a complex scalar field, and

\begin{figure}[h!]
    \centering
    \includegraphics[width=0.3\textwidth]{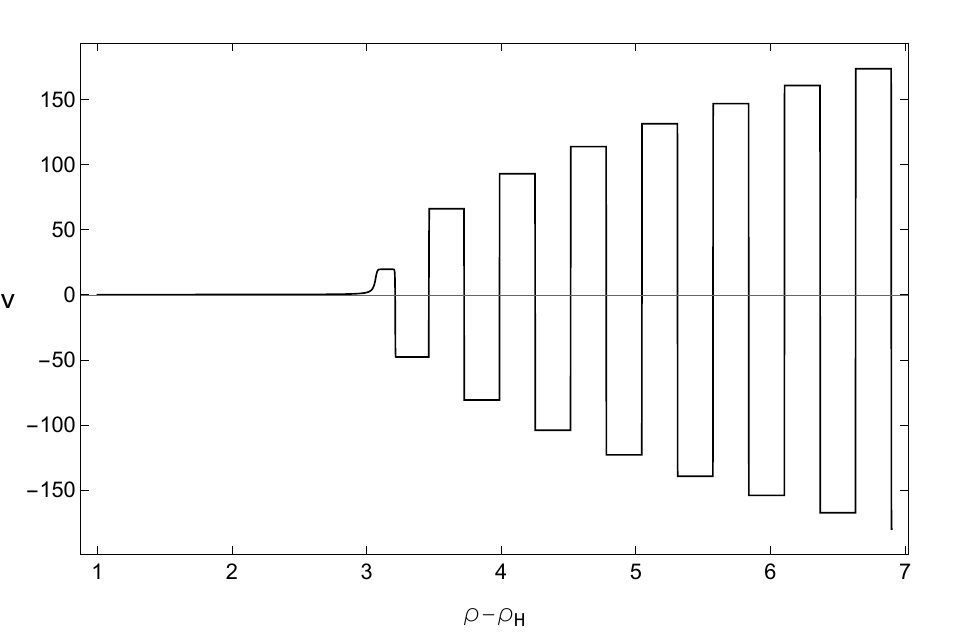}~~
    \includegraphics[width=0.3\textwidth]{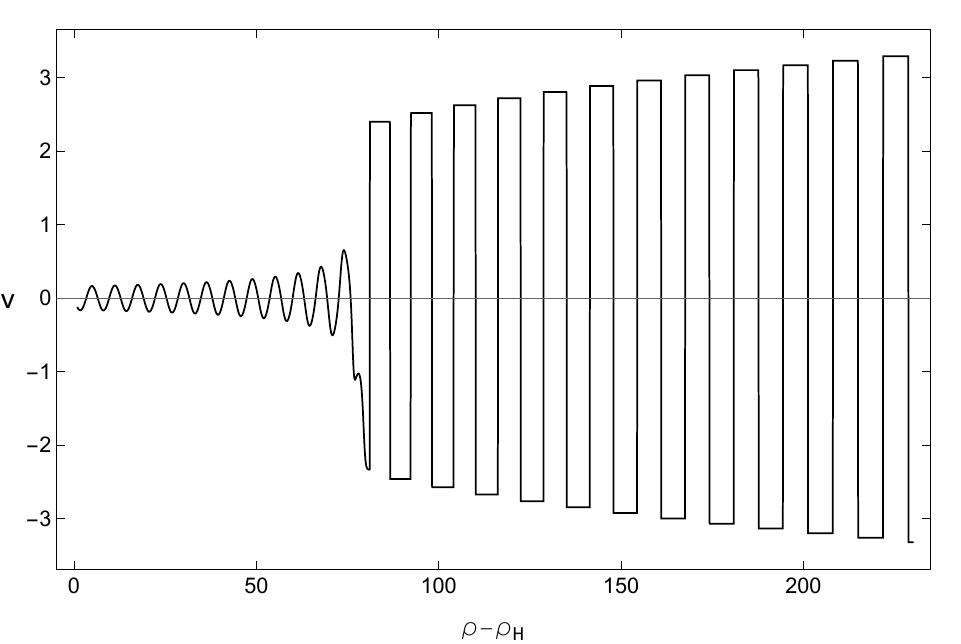}
    ~~
    \includegraphics[width=0.3\textwidth]{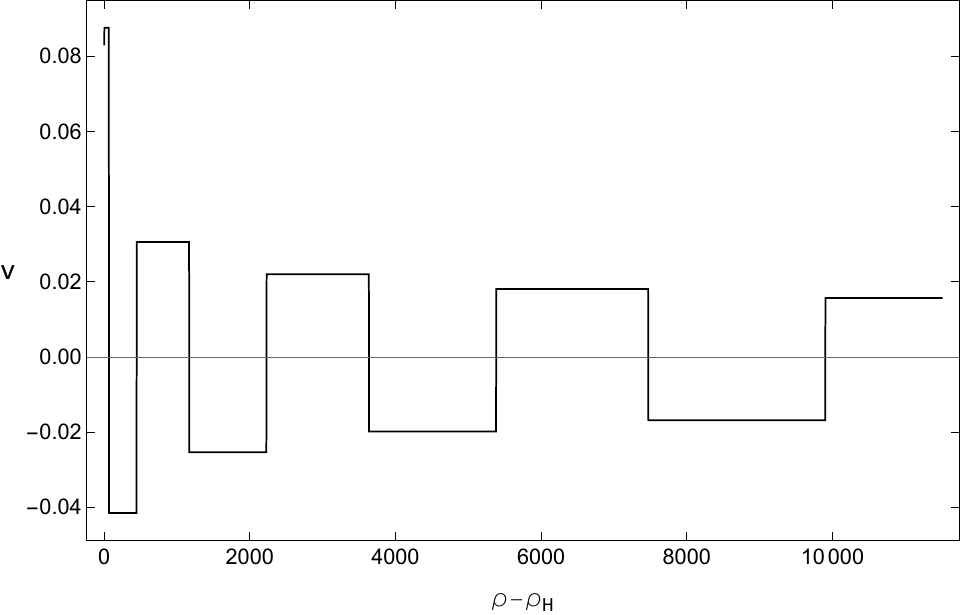}
    \vspace{-0.3cm}
    \caption{\small The evolution of $v$ as a function of $\rho-\rho_H$ in cases (i), (ii) and (iii), from left to right. All of these numerical results are in agreement with \eqref{expr-phiv} and \eqref{expr-nv-lodd}.}
    \label{fig-three-additional-cases}
\end{figure}

(iii) $N=\omega=k=0$: non-rotating black holes with a real scalar field.\\
The corresponding numerical results are shown in Fig. \ref{fig-three-additional-cases}.

Firstly, case (i) differs from the other two. Because of the presence of the rotation, an inversion could occur so that there could exist two types of transitions in the interior, which is expected to be similar to that shown in Fig. \ref{fig-inv-tran}. However, numerically we only find the increasing type of transitions, which is very similar to that shown in Fig. \ref{fig-no-inv-tran}. This indicates an inversion could not exist for case (i), and hence, the initial value of $|v|$ at the first Kasner epoch (i.e. $|v_0|$) 
is always greater than $v_c^T$. This is consistent with our earlier numerical results in \cite{Gao:2023rqc}.  
%although an analytical proof is still lacking. 

{Secondly, since the equations of motion for cases (ii) and (iii) contain no non-negligible rotation terms in the equations of motion, we therefore expect no inversion to occur.} The transitions are of either decreasing or increasing type within a given black hole interior, which depends on whether $|v_0|>v_c^T$. However, we only find the increasing type of transitions in case (ii), while the decreasing type occurs in case (iii). This indicates $|v_0|$ is always greater than $v_c^T$ in case (ii), while $|v_0|<v_c^T$ always holds in case (iii). Therefore, the cases (ii) and (iii) have completely different late-time evolution.

We conclude this section with a puzzle: why the initial value of $|v|$ at the first Kasner epoch is not completely free, but apparently constrained to lie either above or below the critical value $v_c^T$. From the perspective of the equations of motion, no such constraint is apparent. This discrepancy may originate from additional physical restrictions governing the black hole interior, 
%inside the black hole
which is certainly worth further investigation, or it might simply reflect the incompleteness of our numerical survey of the parameter space.

\section{Discussion}

We have presented an analytical study of the interior structure of hairy, rotating black holes in 3D Einstein gravity minimally coupled to a scalar field with a sup-exponential potential. 
Both Kasner inversion and transitions between consecutive Kasner epochs admit analytical solutions. Moreover, we have derived an explicit analytical expression that characterize the infinite sequence of Kasner epochs deep in the interior. 
{During the analytical study, although we have used a specific potential \eqref{expr-potential} to validate the approximations and verify the accuracy of analytical expressions, the analytical method is expected to be valid for the generic potential \eqref{expr-potential-generic} with the conditions \eqref{expr-potential-conditions}. Therefore, the following main physical results are robust for a broad class of potential.}

{The critical value $v_c^I$ for the Kasner inversion and $v_c^T$ for Kasner  transitions play important role in determining the full interior evolution. In particular, these two values are always equal for the generic class of potential \eqref{expr-potential-generic}. This indicates that if a Kasner inversion exists, which is triggered by rotation, it could only affect the fate of interior evolution, namely, change the interior from evolving towards $p_t=0$ and $p_x=1$ to $p_t=1$ and $p_x=0$, without introducing more complicated dynamics.}

At late interior times, despite each individual epoch locally resembling a regular Milne universe on a circle, 
a careful asymptotic analysis of the interplay between  $v\to\infty$ and $\tau\to 0^+$  reveals that the geometry remains a curvature singularity. Therefore, the final nature of this singularity is distinct from that of the rotating BTZ black hole. This contrasts with the four-dimensional, non-rotating case, where the singularity was found to be identical to that of a Schwarzschild black hole \cite{Hartnoll:2022rdv}.

%At late interior times, we show that the geometry still has a curvature singularity from a careful analysis on the order  between $v\to \infty$ and $\tau\to 0$, although each Kasner epoch asymptotically looks like a Milne universe crossed with a circle. Therefore the final nature of the singularity is different from the one in rotating BTZ. This observation is different from the non-rotating black hole in four dimensional case where the singularity is the same as as Schwarzschild black hole \cite{Hartnoll:2022rdv}. 

There are lots of open questions for future study. The analytical description of the Kasner sequence could play a key role in  exploring the holographic dual of black hole interiors. In particular, from the perspective that spacetime geometry emerges from quantum entanglement \cite{Maldacena:2001kr}, it would be highly  interesting to identify the dual CFT structures  responsible for each Kasner epoch and for the infinite cascade of transitions. %sequence of Kasner epochs. 

As we approach the singularity, the spatial volume shrinks to zero. This is the time reverse of a cosmological evolution through an infinite sequence of Kasner epochs near $\tau=0^+$. It would be intriguing to investigate whether this interior dynamics could be reversed to construct new, exactly solvable models of the early universe. 
Furthermore, 
the bouncing structure of Kasner epochs represents a non-chaotic  evolution near the singularity. In contrast, more complicated initial conditions in the BKL picture are expected to give rise to chaotic behavior 
\cite{book, Fleig:2018djg}. 
Constructing a simple chaotic BKL system that can be analytically described would be an interesting challenge. 

While our description assumes the validity of classical gravity, it is well known that quantum effect become important in regions of large curvature. The super-exponential potential might arise from higher dimensional reduction, which itself should be an interesting topic for further study. Realizing this reduction from a UV-complete theory, perhaps through a consistent embedding in string theory, could offer new insights into the quantum nature of the black hole interior. Additionally, exploring the quantization of interior geometry via the Wheeler–DeWitt equation \cite{Hartnoll:2022snh, Caceres:2025xzl} and examining singularity resolution in this context could be useful for our understanding of quantum gravity. 

%%%%%%%%%%%%%%%%%%%%%%%%%%%%%%%%%%%%%%%%%%
%%%%%%%%%%%%%%%%%%%%%%%%%%%%%%%%%%%%%%%%%%
\vspace{.3cm}
\subsection*{Acknowledgments}
We thank Hong-Da Lyu, Ya-Wen Sun, 
You-Jie Zeng, Yu Zhou 
for useful discussions. This work is supported by the National Natural Science Foundation of China grant No. 12375041 and 12575046. 
%%%%%%%%%%%%%%%%%%%%%%%%%%%%%%%%%%%%%%%%%%%%%%%%

\appendix
\section{Details of some expressions  %for the  approximate solutions
}
\label{app:sol}
In this appendix, we show details of some expressions in Sec. \ref{sec:Late-time evolution}. The full solutions for $\chi(\rho)$ and $f(\rho)$ to \eqref{eq:chi-f-rho} are
\be
\begin{split}
\chi(\rho)=&\,c_{\chi }-\frac{2 a_2^2 }{a_3}\tanh \left[a_3 \left(\rho -\rho _n\right)\right]+\frac{\left(a_1-a_2\right){}^2 }{a_3}\log \left[\tanh \left[a_3 \left(\rho -\rho _n\right)\right]+1\right]\\&~~~-\frac{\left(a_1+a_2\right){}^2 }{a_3}\log \left[1-\tanh \left[a_3 \left(\rho -\rho _n\right)\right]\right]\,,\\[0.4pt]
    f(\rho)=&\,c_f\,\frac{ \cosh ^{\frac{2 a_1 a_2}{a_3}+1}\left[a_3 \left(\rho -\rho _n\right)\right] e^{\left(a_1^2+a_2^2+2\right) \left(\rho -\rho _n\right)-\frac{a_2^2 }{a_3}\tanh \left[a_3 \left(\rho -\rho _n\right)\right]-\rho  \left((a_1+a_2 \tanh \left[a_3 (\rho -\rho_n)\right])^2+2\right)}}{a_2 h_n \sinh \left[a_3 \left(\rho -\rho _n\right)\right]+\left(a_1 h_n-1\right) \cosh \left[a_3 \left(\rho -\rho _n\right)\right]}\,,
\end{split}
\ee 
where \(c_\chi\) and \(c_f\) are two constants of integration. From the limit defined in \eqref{eq:chinfkn}, we have 
\be
\begin{split}
    \chi_1^{n}&=c_{\chi }-2 \left(a_1+a_2\right){}^2 \rho _n-\frac{2 a_2 \left(a_2+ 2 a_1 \log 2\right)}{a_3}\,,\\
    \chi_1^{n+1}&=c_{\chi }-2 \left(a_1-a_2\right){}^2 \rho _n+\frac{2 a_2 \left(a_2-2 a_1 \log 2\right)}{a_3}\,,
\end{split}
\ee 
and 
\be
\begin{split}
 f_K^{n}&=c_f\, \frac{4^{-\frac{a_1 a_2}{a_3}} e^{ -\left(\left(a_1+a_2\right){}^2+2\right) \rho _n-\frac{a_2^2}{a_3}}}{a_1 h_n+a_2 h_n-1}\,,\\
    f_K^{n+1}&=c_f\, \frac{4^{-\frac{a_1 a_2}{a_3}}  e^{ -\left(\left(a_1-a_2\right){}^2+2\right) \rho _n+\frac{a_2^2}{a_3}}}{a_1 h_n-a_2 h_n-1}\,.
\end{split}
\ee
Note that between two neighboring epochs during the transitions, we have rewritten the approximate solution \eqref{expr-vappr} for $v$ as
\be 
v_{appr}^T=a_1+a_2 \tanh{\left[a_3(\rho-\rho_n)\right]}\,,
\ee
with
\be 
a_1=\frac{2+c_T}{2h_n}\,,~~~~a_2=\frac{\sqrt{c_T^2-4J_n^2}}{2 h_n}\,,~~~~a_3=\frac{J_n^2 \sqrt{c_T^2-4 J_n^2}}{c_T h_n^2}\,.
\ee 
From these expressions, one obtains the recurrence relations \eqref{eq:relation-chinfn} for \(\chi_1^n\) and \(f_K^n\). 
%%%%%%%%%%%%%%%%%%%%%%%%%%%%%

\end{document}